\documentclass[prb,twocolumn,superscriptaddress]{revtex4}
\usepackage[dvips]{graphicx}
\usepackage{latexsym}
\usepackage{bm}
\begin{document}

\newcommand{\fig}[2]{\includegraphics[width=#1]{#2}}

\title{Ring exchange, the Bose metal, and bosonization in two dimensions}

\author{Arun Paramekanti}
\affiliation{Institute for Theoretical Physics,
University of California, Santa Barbara, CA 93106--4030}
\affiliation{Department of  Physics, University of California,
Santa Barbara, CA 93106--4030}
\author{Leon Balents}
\affiliation{Department of  Physics, University of California,
Santa Barbara, CA 93106--4030}
\author{Matthew P. A. Fisher}
\affiliation{Institute for Theoretical Physics,
University of California, Santa Barbara, CA 93106--4030}

\date{\today}

\begin{abstract}
  Motivated by the high-$T_c$ cuprates, we consider a model of bosonic
  Cooper pairs moving on a square lattice via ring exchange.  We show
  that this model offers a natural middle ground between a
  conventional antiferromagnetic Mott insulator and the fully
  deconfined fractionalized phase which underlies the spin-charge
  separation scenario for high-$T_c$ superconductivity.  We show that
  such ring models sustain a stable critical phase in two dimensions,
  the {\sl Bose metal}.  The Bose metal is a compressible state, with
  gapless but uncondensed boson and ``vortex'' excitations, power-law
  superconducting and charge-ordering correlations, and broad spectral
  functions.  We characterize the Bose metal with the aid of an exact
  plaquette duality transformation, which motivates a universal low
  energy description of the Bose metal.  This description is in terms
  of a pair of dual bosonic phase fields, and is a direct analog of
  the well-known one-dimensional bosonization approach.  We verify the
  validity of the low energy description by numerical simulations of
  the ring model in its exact dual form.  The relevance to the
  high-$T_c$ superconductors and a variety of extensions to other
  systems are discussed, including the bosonization of a two dimensional
  fermionic ring model.
\end{abstract}
\maketitle

\vspace{0.15cm}



\section{Introduction}

Despite intense experimental effort exploring the phase diagram of the
cuprates, the nature of the pseudo-gap regime remains mysterious.  In
the very underdoped normal state there are strong experimental hints
of local pairing and superconducting correlations\cite{pseudo1,pseudo2,pseudo3}, despite the absence
of phase coherent superconductivity.  Within this picture\cite{phasefluc}, the
pseudo-gap regime supports a pair field with non-zero amplitude but with
strong phase fluctuations which disrupt the superconductivity.  A
theoretical approach then requires disordering the superconductivity
by unbinding and proliferating vortices.  Unfortunately, such an
approach will likely face a most worrisome dilemma.  Proliferation and
condensation of single $hc/2e$ vortices necessarily leads to a
confined insulator\cite{BFN2}, which should show sharp features in the electron
spectral function in apparent conflict with ARPES
experiments\cite{pseudo3}.  On the other hand, if {\it pairs} of
$hc/2e$ vortices condense\cite{BFN2,Z2A}, the pseudo-gap phase must necessarily
support gapped unpaired vortex excitations\cite{vison1,vison2,vison3} (``visons'')\cite{Z2B,Z2C}, which have yet
to be observed\cite{visexp1,visexp2}.  Is there an alternative
possibility?  In this paper we find and explore a truly novel quantum
fluid phase of 2d bosons, which we refer to as a ``Bose metal'', which
answers this question in the affirmative.  The Bose metal phase
supports gapless charge excitations, but is {\it not} superconducting
- the Cooper pairs are not condensed.  The $hc/2e$ vortices are
likewise gapless and uncondensed.  Being a stable quantum fluid with
no broken spatial or internal symmetries, the Bose metal phase has
many properties reminiscent of a 2d Fermi liquid - a 1d locus of
gapless excitations (a ``Bose surface''), a finite compressibility, an
almost $T-$linear specific heat with logarithmic corrections.

Neutron scattering measurements in the undoped cuprates have revealed
a zone boundary magnon dispersion which can best be accounted for by
presuming the presence of appreciable four-spin exchange
processes\cite{Aeppli}.  Recent theoretical work points to the
importance of such ring exchange processes driving fractionalization
and spin-charge separation\cite{Z2A,ringnum,BGF}.  Motivated by these
considerations, we focus on a class of model Hamiltonians describing a
square lattice of 2d Bosons (Cooper pairs) with appreciable
ring-exchange.  It is hoped that such microscopic models will be
appropriate in describing the charge sector in the under-doped
cuprates.  When the ring exchange processes are sufficiently strong,
we find that the Bose metal phase is the stable quantum ground state.

Lattice models of interacting bosons in two dimensions, such as the
boson Hubbard model, have been studied quite exhaustively during the
past several decades\cite{supins,subirbook}, primarily as models for Josephson junction
arrays and superconducting films but also in the context of quantum
magnets with an easy plane $U(1)$ symmetry and most recently in the
context of trapped Bose condensates moving in an optical lattice\cite{BEC}.
Typically, the ground state phase diagram consists of a
superconducting phase and one or more insulating states\cite{supins}.  At
fractional boson densities commensurate with the lattice, say at
half-filling, the insulating behavior is driven by a spontaneous
breaking of translational symmetry.  When accessed from the
superconductor, such insulating states can be fruitfully viewed as a
condensation of elementary vortices\cite{MPAFvort}.  Very recent work on a Kagome
lattice boson model with ring exchange\cite{BGF} (which arises in the context of
frustrated quantum magnets) has revealed the existence of a more
exotic fully gapped insulating state with no broken symmetries
whatsoever.  In this state the charged excitations are
``fractionalized'', carrying one-half of the bosons charge\cite{SubirZ2}, and there
are also gapped vortex excitations called visons.  Here, we study the
simpler system of bosons with ring exchange moving on a 2d square
lattice.  Our central finding is the existence of a {\it critical
gapless quantum phase} of $2d$ bosons which is stable over a
particular but generic parameter range.  This phase is in some
respects very similar to the familiar gapless Luttinger liquid phase
of interacting bosons and Fermions moving in one spatial dimension\cite{Lutt1,Lutt2,Lutt3}.
Indeed, to describe this phase we introduce a 2d generalization of
bosonization.  Specifically, we introduce a new duality mapping which
transforms a square lattice model of 2d bosons (in a rotor
representation) with ring exchange, into a 2d theory of ``vortices''
hopping on the sites of the 2d dual lattice.  Then, based on the
symmetries we construct a low energy effective description, in terms
the two (dual) phases of the boson and vortex creation operators.  The
Bose metal phase has a Gaussian fixed point description in terms of
these fields.  In addition, there are various non-quadratic
interactions, which scale to zero in the Bose metal.  Instabilities
towards superconducting and insulating states are triggered when one
or more of these interactions become relevant, closely analogous to
$2k_F$ instabilities in a 1d Luttinger liquid.

Since this paper is quite long, the remainder of the introduction
is devoted to a brief summary of the ring model Hamiltonian
and the bosonized description of the Bose metal.
The Hamiltonian we focus on describes 2d bosons
in a rotor representation:
\begin{equation}
{\cal H}_{\Box}  = {U \over 2} \sum_{\bf r} (n_{\bf r} - \bar{n} )^2
-K \sum_{\bf r} \cos(\Delta_{x y} \phi_{\bf r})  ,
\label{eq:rotorham}
\end{equation}
with
\begin{equation}\label{eq:delxyphi}
  \Delta_{xy} \phi_{\bf r} \equiv
  \phi_{\bf r} - \phi_{{\bf r}+\hat{{\bf x}}}
  - \phi_{{\bf r}+\hat{{\bf y}}}
  + \phi_{{\bf r}+\hat{{\bf x}} + \hat{{\bf y}}}      .
\end{equation}
Here ${\bf r}$ labels sites of the 2d square lattice
and $\phi_{\bf r}$ and $n_{\bf r}$ are canonically conjugate variables:
\begin{equation}
  [\phi_{\bf r}, n_{{\bf r}^\prime} ] = i \delta_{{\bf r}, {\bf r}^\prime }  .
\end{equation}
Representing the phase of the boson wavefunction, $\phi_{\bf r}$ is
taken to be $2\pi$ periodic $\phi_{\bf r} = \phi_{\bf r} + 2\pi$, so
that the eigenvalues of the conjugate boson number operator $n_{\bf
  r}$ are integers.  The mean boson density is set by $\bar{n}$, and
we will primarily focus on the case with half-filling: $\bar{n} = {1
  \over 2}$.  The argument of the cosine in the second term is a
lattice second derivative, involving the four sites around each
elementary square plaquette.  This term hops two bosons on opposite
corners of a plaquette, clockwise (or counterclockwise) around the
plaquette, and is an XY analog of the more familiar $SU(2)$ invariant
4-site ring exchange term for spin one-half operators (which
arises in the context of solid $3-He$\cite{oldring1,oldring2}).

In addition to the conventional spatial, particle-hole, and total
number conservation symmetries
(see Sec.~\ref{sec:effective}), this ring Hamiltonian has an infinite
set of other symmetries.  Specifically, the dynamics of the
Hamiltonian conserves the number of bosons on each row and on each
column of the 2d square lattice - a total of $2L$ associated
symmetries for an $L$ by $L$ system.  This is fewer than a gauge
theory, which has an extensive number of local symmetries, but these
symmetries will nevertheless play a crucial role in constraining the
dynamics of the model and stabilizing a new phase.

We will also consider adding a near neighbor Boson hopping
term,
\begin{equation}
\label{hamhop}
{\cal H}_t = -t \sum_{\langle {\bf r}, {\bf r}^\prime \rangle }
\cos(\phi_{\bf r} - \phi_{{\bf r}^\prime} )  ,
\end{equation}
where the summation is taken over near-neighbor sites on the square
lattice.  This term breaks the $2L$ U(1) symmetries corresponding to
conserved boson numbers on each row and column, leaving only the
globally conserved total boson number.

The remarkable result of this paper is that, over a particular but
generic parameter range, ring models of this type sustain the novel
Bose metal phase. The effective description of the Bose metal is in
terms of low energy ``coarse grained'' variables $\varphi \sim \phi$
and an additional field $\vartheta$ (see Sec.~\ref{sec:effective}),
which is related via $n-\overline{n}\sim \pi^{-1}\Delta_{xy}\vartheta$
to the long-wavelength density, and further can be used to construct a
``vortex creation operator'' $v\sim e^{i\vartheta}$.  To fix notation,
we define the $n_{\bf r},\phi_{\bf r},\varphi_{\bf r}$ operators on
the sites of the original square lattice, with $x,y$ coordinates
taking half-integer values, and $\vartheta_{\bf r}$ (with $N_{\bf
  r},\theta_{\bf r}$ operators to be defined later) on the dual square
lattice with integer $x,y$ coordinates.  The $\varphi,\vartheta$
fields are governed by an approximately Gaussian (Euclidean) effective
action ${\cal S}=\int_0^\beta \!d\tau {\cal L}$, with ${\cal L}={\cal
  L}_0 + {\cal L}_1$, and $\tau$ is imaginary time,
$\beta=1/k_{\scriptscriptstyle B}T$.  The Gaussian part of the
Lagrangian is
\begin{equation}
  {\cal L}_0 = {\cal H}_0[\varphi,\vartheta] + \sum_{\bf r}
  \frac{i}{\pi} \partial_\tau \varphi_{\bf r'}
  \Delta_{xy}\vartheta_{\bf r}, \label{eq:selfdual}
\end{equation}
where the sum is over sites ${\bf r}$, ${\bf r'} = {\bf
  r}+\hat{x}/2+\hat{y}/2$ on the dual and original lattices,
respectively, and the effective Hamiltonian is
\begin{equation}
  {\cal H}_0 = \int_{\bf k} \left[\frac{{\cal
        K}({\bf k})}{2} |(\Delta_{xy} \varphi)_{\bf k}|^2 +
    \frac{{\cal U}({\bf k})}{2\pi^2} |(\Delta_{xy} \vartheta)_{\bf k}|^2
  \right]. \label{eq:effham}
\end{equation}
Here the momentum (${\bf k}$) integral is taken over the Brillouin
zone $|k_x|,|k_y|<\pi$, and $(\Delta_{xy} \varphi)_{\bf k}$ denotes
the Fourier transform of $\Delta_{xy}\varphi_{\bf r}$ (and similarly
for $\Delta_{xy}\vartheta$).  The functions ${\cal K}({\bf k}), {\cal U}({\bf
  k})$ are non-vanishing finite periodic, and analytic.  Their
values along the $k_x$ and $k_y$ axes parameterize the Bose liquid,
much like the effective mass and Fermi liquid parameters do a Fermi
liquid.  Experts will note a strong similarity to the bosonized
effective action for a one dimensional Luttinger liquid\cite{Lutt2,Lutt3}, which is
explored in Sec.~\ref{sec:effective}.  Like in a Luttinger liquid, there are additional
non-quadratic terms in the action.  It is sufficient to keep only
those of the form ${\cal L}_1={\cal L}_{1\varphi}+{\cal
  L}_{1\vartheta}$, with
\begin{eqnarray}
  {\cal L}_{1\vartheta} & = & \sum_{\bf r} \sum_{q,{\bf s}} t_q^{m,n}
  \cos[q(\varphi_{\bf r}\!-\!\varphi_{\bf 
    r+s})], \label{eq:L1p}
\end{eqnarray}
with ${\bf r}$ on the original lattice, and
\begin{eqnarray}
  {\cal L}_{1\varphi} & = & -\sum_{\bf r} \bigg\{ \sum_{q=1}^\infty
  \upsilon_{2q} \cos 
  (2q\vartheta_{\bf r}+qQ xy) \nonumber \\ 
  & & \hspace{-0.7in} +
  \!\sum_{q{\bf s}} w_{2q}^{m,n}\cos
  [2q(\vartheta_{\bf r}\!-\!\vartheta_{\bf
    r+s})\!-\!qQ(nx\!+\!my\!+\!mn)]\bigg\}, \label{eq:L1t}
\end{eqnarray}
with ${\bf r}$ on the dual lattice.
In both Eqs.~(\ref{eq:L1p}-\ref{eq:L1t}), $Q=2\pi\bar{n}$, ${\bf
  r}=(x,y)$, ${\bf s}=(m,n)$.  Here, $w_{2q}^{m,n}=w_{2q}^{n,m}$ gives
the amplitude to hop $2q$ vortices by the translation vector $(m,n)$
(or $(n,m)$), and $t_q^{m,n}=t_q^{n,m}$ gives the hopping amplitude
for a $q$ bosons along the same vector.  In the Bose metal phase, all
these terms are ``irrelevant'', i.e. give only perturbative
corrections to physical properties, though these corrections can be
significant.  Also like in a Luttinger liquid, there is a non-trivial
relation between microscopic quantities such as the energy and
particle densities and the coarse grained variables of the low energy
theory.  In particular, for the energy and particle densities, one
finds
\begin{eqnarray}\label{eq:opcontentn}
\delta n_{x+\frac{1}{2},y+\frac{1}{2}} & \sim & c_0^\rho
\Delta_{xy}\vartheta_{xy}
 \nonumber \\
& &   + \sum_{q=1}^\infty c_{2q}^\rho\Delta_{xy} \sin(
2q\vartheta_{xy} \!+\!
qQxy) , \\
\varepsilon_{xy} & \sim & c_0^\varepsilon\dot{\vartheta}_{xy}^2 +
\sum_{q=1}^\infty c_{2q}^\varepsilon \cos( 2q\vartheta_{xy}+qQxy),
\label{eq:opcontente} 
\end{eqnarray}
where $\delta n = n-\bar{n}$, $\varepsilon$ is the energy density, and
$c_{2q}^{\rho/\varepsilon}$ are non-universal constants.

The paper is organized as follows.  In Section 2 we 
treat the ring exchange term within a ``spin-wave'' approximation.
This reduces the Hamiltonian to a quadratic form which can
be readily diagonalized.
Within the two-dimensional Brillouin
zone there are gapless excitations along the lines $k_x=0$ and $k_y=0$,
which are associated with an
infinite set of conservation laws possessed by the
ring exchange model.  
Dual ``vortex'' variables are then introduced
via a new plaquette duality transformation.
This dual representation is well suited
to numerical simulations since it is free of any ``sign problems''.
We implement a quantum Monte Carlo simulation, and show that the Bose
metal phase is present over appreciable regions of the phase diagram
of the ring-exchange rotor model. 

In Section III we construct the low energy effective model
in terms of the dual vortex fields, and extract the universal
properties of the Bose metal phase in Section IV.
These properties are confirmed
by the quantum Monte Carlo results.
Section  V is devoted to an analysis of the stability of the Bose metal phase.  
Remarkably, we find that for generic (incommensurate) boson densities,
there are regions of parameters where the Bose metal phase is stable towards all perturbations, even those that break the row/column symmetries.
The situation is reminiscent
of the Fermi liquid phase, whose ``fixed point'' description
possesses an enormous set of ``emergent symmetries'' - the number
of Fermions on each patch of the Fermi surface are
independently conserved.
At commensurate fillings we explore in some detail the
instabilities of the Bose metal towards various insulating states.

In Section 6 we consider a ``hard-core'' version of the boson Hamiltonian
which allows us to obtain an exact zero energy wavefunction
when the couplings are carefully tuned.
We then perturb away from the soluble point,
and compute properties of the adjacent quantum phase
which we thereby identify as the Bose metal.
Finally, in Section 7 we conclude with a discussion
of how various ring exchange processes
emerge from a $Z_2$ gauge theory formulation
of interacting electrons and then make some speculative
remarks about the possible relevance of the Bose metal phase
in the context of the under-doped cuprates.

\section{\label{sec:Duality} Spin Waves and Plaquette Duality}

\subsection{Spin Waves, massless modes and symmetry}

Here we consider a spin wave approximation to the
microscopic ring Hamiltonian which leads to a harmonic and soluble theory.
The resulting energy spectrum vanishes along two lines in
momentum space.  This remarkable feature is then shown to follow directly
from the existence of the infinite set of conservation laws
of the ring Hamiltonian.  As we detail in later sections,
such a harmonic description underlies an effective theory
of the Bose-metal phase, in close analogy to bosonization
of the 1d Luttinger liquid.

To this end,
it is useful to employ a
combination of path integral and Hamiltonian methods.  A standard path
integral representation is constructed in the usual way using $\phi$
eigenstates.  In the time continuum limit, the partition function
for the pure ring Hamiltonian takes the form,
\begin{equation}
  Z= {\rm Tr} e^{-\beta H_{\Box}} = \int [d\phi_{{\bf r}\tau}]
  e^{-\int_0^\beta \! d\tau L^\phi},
\end{equation}
with the Lagrangian
\begin{equation}
  L^\phi = \sum_{\bf r} [\frac{1}{2U} (\partial_\tau\phi_{\bf r})^2 + i
  \overline{n}\partial_\tau\phi_{\bf r} - K \cos(\Delta_{xy}\phi_{\bf
    r})] . \label{eq:philag}
\end{equation}
The resemblance of the above Lagrangian with that for
the standard Boson-Hubbard model\cite{supins} (which has a single, rather than
double, lattice derivative inside the cosine) suggests
that one might try expanding the cosine potential
to quadratic order.  Doing so gives a soluble Harmonic theory
for the action, which can be readily diagonalized as,
\begin{equation}
{\cal S}_{\rm spin wave} = {1 \over 2U} \int { d^2 k \over (2\pi)^2} \int_{-\infty}^{\infty}  {d \omega \over 2\pi} [\omega^2 +
E_{\bf k}^2 ]
| \phi({\bf k},\omega) |^2    ,
\end{equation}
with ${\bf k} = (k_x,k_y)$ living in the first Brillouin zone,
$k_x,k_y \in [-\pi, \pi]$.
Remarkably, the energy dispersion,
\begin{equation}
E_{\bf k} = 4 \sqrt{UK} |\sin(k_x/2) \sin(k_y/2) | ,
\end{equation}
vanishes on both the $k_x$ and $k_y$ axis.  The presence of these
gapless excitations can be traced directly to the existence of the
infinite set of symmetries of the ring Hamiltonian, which conserves
the number of bosons on each row and on each column of the 2d square
lattice.  Specifically, these symmetries imply an invariance of the
energy (and action) under,
\begin{equation}
\phi_{\bf r} \rightarrow \phi_{\bf r} + \Phi_x(x) + \Phi_y(y)   ,
\end{equation}
for arbitrary functions $\Phi_x(x)$ and $\Phi_y(y)$.
This invariance dictates that in the Harmonic spin wave form
the energy must vanish whenever $k_x=0$ or $k_y =0$.

Since the plaquette term involves a lattice second derivative rather
than an ordinary lattice gradient, however, it is clear that phase
($\phi$) fluctuations will be large in the ring model, and the
spin wave expansion is suspect.  Indeed, even in the classical limit
$U\rightarrow 0$, one can readily see that ``vortex'' configurations
in which $\phi_{\bf r}$ winds by $2\pi$ around some plaquette ({\sl
  e.g.} for a vortex with center at $x=y=0$, $\phi_{xy} =
\pi/2\Theta(-x)\Theta(y)+\pi\Theta(-x)\Theta(-y)+
3\pi/2\Theta(x)\Theta(-y)$, where $\Theta(z)$ is the Heavyside step
function) are {\sl finite} in energy rather than logarithmic as in an
ordinary XY model.  Further, ``double vortex'' configurations in which
this winding is $4\pi$ can be smoothly deformed into zero energy
conformations (e.g. $\phi_{xy}=\pi\Theta(-x)+\pi\Theta(-y)$).  This
suggests that for non-zero $U$ there will be substantial ``vorticity''
in the low-lying states.

To address the legitimacy of the spin wave expansion,
it is necessary to account for the periodicity of the cosine potential.
This is most readily accessed by transforming to a dual form,
just as one transforms to a dual bosonized representation
for a system of 1d particles\cite{Lutt3}.
Here we introduce a new 2d quantum duality transformation\cite{oldduality}
which is specifically tailored for the ring model.

\subsection{Plaquette Duality}

We consider dual fields living on the sites of the dual lattice,
denoted $\theta_{\bf r}$ and $N_{\bf r}$, which are canonically
conjugate variables:
\begin{equation}
[N_{\bf r}, \theta_{{\bf r}^\prime} ] = i \delta_{{\bf r}, {\bf
    r}^\prime }  .
\end{equation}
Here we take $\pi N_{\bf r}$ to be $2\pi$ periodic so that
$\theta_{\bf r} /\pi$ has integer eigenvalues.  The reason for this
unusual choice of normalization should become apparent below.  For
notational ease we are denoting the sites of both the original and
dual square lattices as ${\bf r}$.  In analogy to
Eq.~(\ref{eq:delxyphi}), it is convenient to introduce an operator
defined on the plaquettes of the dual lattice (i.e. sites of the
original lattice) as:
\begin{equation}
  \Delta_{xy} \theta_{\bf r} \equiv
  \theta_{\bf r} - \theta_{{\bf r}+\hat{{\bf x}}}
  - \theta_{{\bf r}+\hat{{\bf y}}}
  + \theta_{{\bf r}+\hat{{\bf x}} + \hat{{\bf y}}}      .
\end{equation}
These two new dual fields are related to the original fields by the
relations,
\begin{equation}
  \pi N_{\bf r^\prime} = \Delta_{xy} \phi_{\bf r}   ;   \hskip1cm
 \Delta_{xy} \theta_{\bf r} = \pi n_{\bf r^\prime}, \label{eq:dualvars}
\end{equation}
where ${\bf r}^\prime = {\bf r} + \hat{x}/2 + \hat{y}/2$.
One can check that with $N_{\bf r}$ and $\theta_{\bf r}$ as conjugate
fields, the original variables satisfy the required commutation
relations.  

To interpret the new variables, consider now a vortex
centered on some site $(x,y)$ of the dual lattice (the ``core'').
Classically, for the four sites on plaquette of the original lattice
surrounding this site,
\begin{equation}
e^{i\phi_{x+a/2,y+b/2}} =  \left(\frac{a+ib}{\sqrt{2}}\right)^{N_v},
\end{equation}
where $a,b=\pm 1$, and $N_v$ denotes the number of vortices
(vorticity) on this plaquette.  Comparing to Eq.~(\ref{eq:dualvars}),
one finds $N=N_v \, {\rm (mod\,2)}$, so $N$ can be interpreted as a
vortex number operator, modulo two.  Since $\theta$ and $N$ are
canonically conjugate, the operators $e^{\mp i\theta}$ perform canonical
``translations'' of $N$, and hence can be regarded as vortex creation
and annihilation operators, respectively.  Note that since $\theta \in
\pi {\cal Z}$, $e^{2i\theta} = 1$, consistent with the ambiguity in
$N=N_v$ under even integer shifts.  Clearly periodicity of $\phi$ in
the original variables is encoded in the discreteness of $\theta$
in the dual description.

The plaquette Hamiltonian when re-expressed in the dual variables reads,
\begin{equation}
H_{\Box} = -K \sum_{\bf r} \cos(\pi N_{\bf r}) + {U \over
  2\pi^2} \sum_{\bf r} [\Delta_{xy} \theta_{\bf r}
- \pi\bar{n} ]^2   .
\end{equation}
We will for the time being neglect the tunneling term $H_t$,
which will be returned to later.  Note the strong similarity between
the dual and original forms.  To bring this out more clearly, and for
the numerical simulations to be considered shortly, it is useful to go
to a path integral formulation.  In particular, consider the
partition function
\begin{equation}
  Z_\Box = {\rm Tr}\, e^{-\beta H_\Box}.
\end{equation}
Expanding $Z_\Box$ in the usual Trotter fashion with a time-slice
$\Delta\tau=\epsilon \rightarrow 0^+$, using the discrete basis of
eigenstates of $\theta_{\bf r}$, one finds (Appendix~\ref{app:trotter})
\begin{equation}
  Z = \sum_{\{ \theta_{\bf r}(\tau) \}} e^{ - \left( \epsilon
      \sum_{\tau} \right)
    L_\Box}, \label{eq:Zdiscrete}
\end{equation}
with the ``Lagrangian''
\begin{equation}
  L_\Box[\theta] = \sum_{\bf r}\bigg\{\frac{\epsilon}{\pi^2}\ln(\frac{2}{\epsilon K})
  (\partial_\tau \theta_{\bf r})^2   + {U \over 2\pi^2}
  [ \Delta_{xy} \theta_{\bf r} - \pi\overline{n}  ]^2  \bigg\} ,
  \label{eq:thetalag}
\end{equation}
where $\partial_\tau \theta_{\bf r} = [\theta_{\bf
    r}(\tau+\epsilon)-\theta_{\bf r}(\tau)]/\epsilon$.
The $\epsilon$-dependence of the time derivative term in
Eq.~(\ref{eq:thetalag}) is familiar from the ``time-continuum
limit'' relating e.g. the $d+1$-dimensional classical and
$d$-dimensional quantum transverse-field Ising models.  Note the
strong similarity of Eq.~(\ref{eq:thetalag}) to Eq.~(\ref{eq:philag}),
which emphasizes their nearly self-dual nature.

\begin{figure}
\begin{center}
\vskip-2mm
\hspace*{0mm}
\centerline{\fig{2.8in}{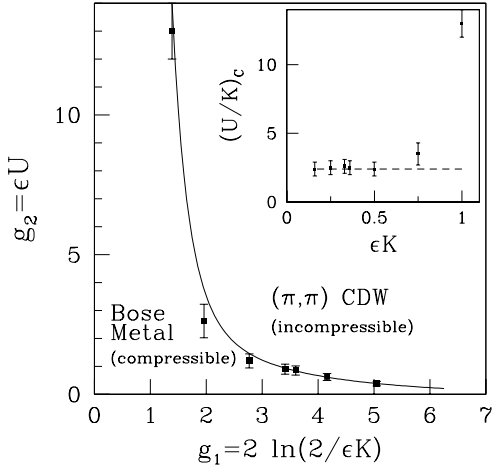}}
\vskip-2mm
\caption{The phase diagram of the time discretized dual model
as a function of the two coupling constants,
showing a compressible Bose metal phase and an incompressible
phase which we identify as a $(\pi,\pi)$ charge density wave.
The compressibility $\kappa$ was obtained from the ${\bf k} \to 0$
extrapolation of 
the density susceptibility $\chi_{n n}({\bf k},\omega=0)$ on 
$8 \times 8\times 32$ lattices as indicated, with some checks 
made on larger system sizes.
As seen from the inset, the critical $U/K$ stays finite in the
time continuum limit $\epsilon K \to 0$ indicating a phase 
transition at $U/K \approx 2.4 \pm 0.4$ in the quantum model.}
\label{fig:phasediag}
\end{center}
\end{figure}

The formulation in Eqs.~(\ref{eq:Zdiscrete}-\ref{eq:thetalag}) is
quite convenient for numerical simulations.  
For the simulations, we define an integer-valued ``height'' field 
$\tilde\theta_{\bf r}(\tau)$ through
\begin{equation}
\Delta_{xy}\theta_{\bf r}(\tau) = \pi n^{(B)}_{{\bf r}} + \pi
\Delta_{xy} \tilde\theta_{\bf r}(\tau)
\end{equation}
where $n^{(B)}_{{\bf r}} = [1+(-1)^{x+y}]/2$ is the background 
staggered density, chosen such that $\tilde\theta_{\bf r}(\tau)$ obeys 
periodic boundary conditions. In this language, we obtain a
generalized (anisotropic) solid-on-solid model in $2+1$ 
dimensions. The correlation functions for the $\theta$-fields are
easily re-expressed (and numerically evaluated) in terms of these
height variables, using Monte Carlo methods. 
An indication of the sort of results obtained in shown in Fig.~\ref{fig:phasediag}
which presents the phase diagram of this model 
as a function of $U/K$ and $\epsilon$, based on an evaluation 
of the density correlations.
For the simulations, we worked on a $L_x\times L_x \times L_\tau$
lattice, with various system sizes indicated in the paper. We
used a Metropolis algorithm with a single-site update, $\tilde\theta_{\bf r}(\tau) 
\to\tilde\theta_{\bf r}(\tau) \pm 1$, which corresponds to a ring-exchange
move for the boson density. We checked for equilibration of
various quantities, and averaged the data over $10^4-10^6$
sweeps of the lattice depending on the correlation function and 
location in the phase diagram.
Notice that in the time continuum limit ($\epsilon \rightarrow 0$),
the simulations reveal two phases as the dimensionless ratio $K/U$
is varied, separating an ordered charge density wave state at large $U$
from the Bose metal phase when the ring exchange term is large.

Before obtaining an effective low-energy description,
it is convenient to rewrite the dual partition function,
Eq.~(\ref{eq:Zdiscrete}) in terms of {\sl continuous} $\theta$
variables using the Poisson summation formula,
\begin{equation}
  Z = \int [d\theta_{\bf r}(\tau)] \sum_{s_{\bf r}(\tau)}
  e^{\sum_{\bf r}(\tau) ( 2 i s_{\bf r}(\tau) \theta_{\bf
        r}(\tau)-\delta |s_{\bf r}(\tau)|) } e^{
    -\epsilon \sum_{\tau} L_\Box [\theta]}. \label{eq:poisson1}
\end{equation}
In Eq.~(\ref{eq:poisson1}), we have included a parameter $\delta \ll
1$, which ``softens'' the discrete-$\theta$ constraint (exactly
imposed for $\delta=0$).  Carrying out the sum over the integer-valued
$s_{\bf r}(\tau)$ variables, one finds
\begin{equation}
  Z = Z_0 \int [d\theta_{\bf r}(\tau)] e^{- S_0 - S_1}, \label{eq:poisson2}
\end{equation}
where $Z_0$ is a constant (divergent for $\delta\rightarrow 0$),
\begin{eqnarray}
S_0 & = & \sum_{\tau}
    \epsilon L_\Box[\theta], \\
S_1 & = &  - \sum_{{\bf r}\tau} \sum_{q=1}^\infty v_{2q} \cos
    2q\theta_{\bf r}(\tau)
\end{eqnarray}
and the $v_{2q}$ are O(1) coefficients whose precise values are not
important.  It is convenient to drop the constant and introduce
infinitesimal source
fields $h^\rho$, $h^\varepsilon$, for the density and energy density,
respectively,
\begin{equation}
  Z[h^\rho,h^\varepsilon] = \int [d\theta_{\bf r}(\tau)] e^{-
    S_0 - S_1 - S_h},\label{eq:poisson3}
\end{equation}
with
\begin{equation}
S_h= -\sum_{{\bf r},\tau} \left[h^\rho_{\bf r'}(\tau)
      \left(\Delta_{xy}\theta_{\bf r}-\pi \overline{n}\right)+
      h^\varepsilon_{\bf r}(\tau)
      (\partial_\tau\theta_{\bf r})^2\right],
\end{equation}
where ${\bf r}^\prime = {\bf r} + \hat{x}/2 + \hat{y}/2$.

\section{\label{sec:effective}Effective Model}

We now make an educated but perhaps bold guess as to the nature (and
existence!) of a low energy effective description.  In the spirit of the
renormalization group and effective field theory, we imagine defining
a {\sl temporally} coarse-grained variable $\vartheta$ in which the
high-frequency modes of $\theta$ are averaged over
\begin{equation}
  \vartheta_{\bf r}(\tau)= \left[\theta_{\bf r}(\tau) \right]_f -\pi\overline{n}xy,
 \label{eq:varthetadef}
\end{equation}
where the square brackets indicate an average over ``fast''
high-frequency modes, and we have for convenience removed the mean
``curvature'' in $\vartheta$ by a non-fluctuating spatially dependent
shift.  Provided that we average only over high-frequency modes, we
expect that the resulting effective description of $\vartheta$ will
remain local both in space and (imaginary) time.  Inspection of
Eq.~(\ref{eq:poisson2}) shows that the dual microscopic action for
$\theta$ can be written as a sum of a quadratic part ($\sum_{\tau,{\bf
    r}} \epsilon {\cal L}_\Box$) and non-quadratic corrections.  We
postulate that the low-energy effective action for $\vartheta$ is a
sum of a renormalized quadratic term similar to ${\cal L}_\Box$ and
{\sl small} non-quadratic corrections.  Although we have not
implemented a detailed renormalization group treatment, the latter
statement is tantamount to declaring the system is controlled by a
stable Gaussian fixed point.  Obviously such a fixed point can have at
best a large finite basin of attraction, and we will return to the
problem of determining the region of stability of the Gaussian theory
in Sec.~\ref{sec:instab}.

\subsection{Symmetries}

Formally, the above postulate of the effective field theory
description can be formulated as the statement that the generating
functional can be approximated for low-frequency source fields by
$Z[h^\rho,h^\varepsilon] \approx {\cal Z}_0 {\cal
  Z}[h^\rho,h^\varepsilon]$, where ${\cal Z}_0$ is an irrelevant
constant, and
\begin{equation}
  {\cal Z} = \int [d\vartheta_{\bf r}(\tau)] e^{-{\cal S}_0[\vartheta]
    - {\cal S}_1[\vartheta] - {\cal S}_h[\vartheta]},
\end{equation}
where ${\cal S}_0[\vartheta]$ is a renormalized quadratic form, ${\cal
  S}_1$ contains small renormalized non-quadratic perturbations, and
${\cal S}_h$ is linear in $h^\rho$, $h^\varepsilon$ (we drop higher
order terms in the infinitesimal sources).  In general, we
expect all three parts of the action to take the most general
form possible consistent with locality, analyticity (since only
high-frequency modes are averaged over), and the symmetries of the
problem, which should be found from the microscopic partition
function, Eq.~(\ref{eq:poisson3}).  These are
\begin{itemize}
\item Row/Column symmetry: $\theta_{x,y}(\tau) \rightarrow
  \theta_{x,y}(\tau) + \pi [ m_x(x) + m_y(y) ]$, where $m_x(x)$ and
  $m_y(y)$ are {\sl arbitrary} integer-valued functions.  This
  corresponds to conservation of the number of vortices (modulo 2) on
  each row and column, 
  a dual version of the conserved boson number
  on each row/column of the original lattice.
\item Space and time translations: $\theta_{\bf r}(\tau)
  \rightarrow \theta_{\bf r+R}(\tau+\tau_0)$, where ${\bf R}$ is an
  arbitrary lattice vector and $\tau_0$ is an arbitrary real number.
  Under this translation, the sources must also be translated, $h_{\bf
    r}^{\rho/\varepsilon}(\tau) \rightarrow h_{\bf
    r+R}^{\rho/\varepsilon}(\tau+\tau_0)$.
\item Reflections across a row or column containing a site of the dual
  lattice (or bonds of the original lattice): $\theta_{x,y} \rightarrow s
  s'\theta_{sx,s'y}$, where $s,s'=\pm 1$.  The sources transform as
  $h_{x,y}^{\rho} \rightarrow s s' h^\rho_{s x,s' y}$,
  $h_{x,y}^\varepsilon \rightarrow h^\varepsilon_{sx,s'y}$.
\item Reflections across a row or column containing bonds of the dual
  lattice (or sites of the original lattice): This is not independent,
  and can be obtained from a composition of a translation and a
  site-reflection.  But for completeness, $\theta_{x,y} \rightarrow
  s s' \theta_{1/2+s(x-1/2), 1/2+s'(y-1/2)}$, $h_{x,y}^{\rho}
  \rightarrow s s' h^\rho_{1/2+s(x-1/2), 1/2+s'(y-1/2)}$,
  $h_{x,y}^\varepsilon \rightarrow h^\varepsilon_{1/2+s(x-1/2),
    1/2+s'(y-1/2)}$.
\item Time-reversal: $\theta_{\bf r}(\tau) \rightarrow
  \theta_{\bf r}(-\tau)$, $h_{\bf
    r}^{\rho/\varepsilon}(\tau) \rightarrow h_{\bf
    r}^{\rho/\varepsilon}(-\tau)$.
\item Four-fold rotations: $\theta_{x,y} \rightarrow \theta_{y,-x}$,
  $h^\rho_{x+1/2,y+1/2} \rightarrow - h^\rho_{y+1/2, -x-1/2}$,
  $h^\varepsilon_{x,y}\rightarrow h^\varepsilon_{y,-x}$.
\item Particle-hole symmetry: This is an invariance {\sl only} for
  integer or half-integer densities ($2\overline{n} \in {\cal Z}$).
  For such values, the symmetry operation is $\theta_{xy} \rightarrow
  2\pi \overline{n} xy - \theta_{xy}$, $h^\rho_{\bf r} \rightarrow -
  h^\rho_{\bf r}$, $h^\varepsilon_{\bf r} \rightarrow
  h^\varepsilon_{\bf r}$.
\end{itemize}

\subsection{Gaussian action}

The form of ${\cal S}$ is dictated by these symmetries and the
relation between the microscopic and coarse-grained fields,
Eq.~(\ref{eq:varthetadef}).  Note that the shift by
$\pi\overline{n}xy$ implies that $\vartheta$ is {\sl not} a
scalar.  Consider first ${\cal S}_0$. By space and time
translational symmetry, it is diagonal in momentum and frequency
space,
\begin{equation}
\label{Gaussian}
  {\cal S}_0 = {1 \over 2} \int_{\bf k}\int_\omega{\cal M}({\bf k},
  \omega)  | \vartheta({\bf k},\omega) |^2,
\end{equation}
where $\int_{\bf k} \equiv \int \! d^2{\bf k}/(2\pi)^2$,
$\int_\omega \equiv \int_{-\infty}^\infty \! d\omega/(2\pi)$, and the
momentum integral is taken over the Brillouin zone $|k_x|,|k_y|<\pi$.
By row/column translational symmetry,
\begin{equation}
  {\cal M}(k_x=0,k_y,\omega = 0)  = 0   \hskip0.5cm {\rm for \, all\,} k_y ,
\end{equation}
and similarly for $k_x \leftrightarrow k_y$.  The latter condition
implies the existence of {\sl gapless} excitations along the $k_x$ and
$k_y$ axes in momentum space.  These are the only such gapless states
mandated by symmetry, and we expect the low energy physics to be
dominated therefore by momenta near these axes and by low frequencies.
Quite generally, the kernel ${\cal M}$ can be expanded near the zero
modes at $k_x=\omega=0$, and at lowest order takes the form,
\begin{equation}
  {\cal M}({\bf k},\omega) = A(k_y)  \omega^2 + B(k_y) k_x^2  , \label{eq:expand1}
\end{equation}
where the expansion functions $A(k_y)$ and $B(k_y)$
are even and $2 \pi$-periodic in $k_y$.
A similar expansion is of course possible around
$k_y=\omega=0$, with the {\it identical} expansion coefficients
due to rotational symmetry.  Analyticity at ${\bf k}=0$ further implies $B(0)=0$.
A convenient representation of ${\cal M}$ which satisfies these
requirements is
\begin{equation}
\label{kernel}
  {\cal M}({\bf k},\omega) = \frac{\omega^2 + E_{\bf k}^2}{\pi^2{\cal K}({\bf k})},
\end{equation}
with mode energy,
\begin{equation}
\label{modeenergy}
E_{\bf k} \equiv 4 \sqrt{{\cal K}({\bf k}) {\cal U}({\bf k})}|\sin(k_x/2) \sin(k_y/2)|   .
\end{equation}
Here, ${\cal U}({\bf k})$ and ${\cal K}({\bf k})$
are positive $2\pi$ periodic functions which characterize
the Bose metal phase, but it is only their
behavior for $k_x \ll 1$ or $k_y \ll 1$ that determine the 
universal low-energy properties of the
theory.

\subsection{Interaction terms}

Consider next the interaction terms, ${\cal S}_1=\int\!d\tau\, {\cal L}_1$.  Locality
requires that they couple combinations of $\vartheta_{\bf r}$
fields only with nearby points ${\bf r}$. Hence we consider a
successive sequence of terms, ${\cal L}_1 = \sum_{m} {\cal
L}_{1;m}$, coupling $\vartheta$ fields at a total of $m$ distinct
points. We expect that the ${\cal L}_{1;m}$ becomes
(exponentially) increasingly small with increasing $m$. First
consider the single-site terms, which are highly constrained,
particularly by translational invariance, under which
$\vartheta_{xy}+\pi\overline{n}xy$ transforms as a scalar.
Generally,
\begin{equation}\label{eq:onsite}
  {\cal L}_{1;1} = -\sum_{xy} \left[\sum_{q=1}^\infty \upsilon_{2q} \cos
  (2q\vartheta+qQ xy) + \sum_{q=2}^\infty {\cal K}_q^{-1}
  (\partial_\tau \vartheta)^{2q} \right],
\end{equation}
where $Q=2\pi\bar{n}$, and has the physical meaning of the
smallest reciprocal lattice vector of a one dimensional lattice
with density $\bar{n}$ (like $2k_F$ for a charge density wave).
The second set of terms involve more time-derivatives than the
analogous quadratic interactions in ${\cal S}_0$, and hence are
negligible at low frequency, so we will take ${\cal K}_q^{-1}=0$
in the following. The $\upsilon_{2q}$ terms will play an important
role in Sec.~\ref{sec:instab}.  Note that, except when the density
takes special commensurate values (such as the interesting case
$\overline{n}=1/2$), they are strongly oscillatory in space.  

Next consider two-site terms.  Similar to the $K_q^{-1}$ terms in
Eq.~(\ref{eq:onsite}), a variety of spatial (lattice) and time
derivative invariants are possible, but are negligible relative to
${\cal S}_0$, so we do not include them here.  More interesting
are sinusoidal terms, which must take the form
\begin{widetext}
\begin{eqnarray}\label{eq:twosite}
  {\cal L}_{1;2} & = & -\sum_{xy;m,n} \sum_{qq'} \bigg\{
  w_{2q,2q',mn}^- \cos \left[2q\vartheta_{xy}-2q'\vartheta_{x+m,y+n} +
  Q(q x y-q' (x+m)(y+n))\right] \nonumber \\
  & & + w_{2q,2q',mn}^+ \cos \left[2q\vartheta_{xy}+2q'\vartheta_{x+m,y+n} +
  Q(q x y+q' (x+m)(y+n))\right]\bigg\},
\end{eqnarray}
\end{widetext}
where rotational and reflection invariance require
$w_{2q,2q',nm}^\pm = w_{2q,2q',mn}^\pm$, and without loss of
generality we may also impose $w_{2q',2q,nm}^\pm =
w_{2q,2q',mn}^\pm$.  Like the $\upsilon_{2q}$ terms in
Eq.~(\ref{eq:onsite}), most of the operators in
Eq.~(\ref{eq:twosite}) are highly oscillatory for generic
$\bar{n}$.  An important exception arises in $w_{2q,2q,mn}^-$ for
rational densities $\bar{n}=z'/z$, where $z,z'$ are integers. Then
these terms are non-oscillatory if $z' m q$ and $z' n q$ are $z$
times integers.  We consider in particular the case $z'=1$, for
which $m,q$ and $n,q$ can be chosen as all possible factors of $z \ell$
into two integers, with arbitrary integer $\ell$.  Keeping only these terms, one has
\begin{eqnarray}\label{eq:twosite2}
  \!\!\!\!\!\!\!\!{\cal L}_{1;2} & \!\! = \!\! & -\sum_{xy} \sum_{m\cdot q=n\cdot q=z\ell} \!\!\!\!\!\!\!w_{2q}^{m,n} \cos
  [2q(\vartheta_{xy}-\vartheta_{x+m,y+n})],
\end{eqnarray}
where, $w_{2q}^{m,n}=w_{2q}^{n,m}$ by rotational invariance.  Since
$e^{i\vartheta}$ acts as a vortex creation operator, the terms in
Eq.~(\ref{eq:twosite2}) can be thought of as hopping $2q$ vortices a
distance ${\bf s}=(m,n)$ on the 2d dual lattice.

Finally, consider ${\cal S}_h=\int\!d\tau\,{\cal L}_h$.  As for ${\cal
  L}_1$, we expect that the largest contributions will occur in terms
involving $\vartheta_{\bf r}$ fields at a small number of distinct
points ${\bf r}$.  For simplicity, we keep only the most local of
these.  Employing the symmetries above, one finds
\begin{eqnarray}\label{eq:sh}
{\cal L}_h & = & \sum_{xy} \bigg\{
h^\rho_{x+1/2,y+1/2}\bigg[c_0^\rho \Delta_{xy}\vartheta_{xy}
 \nonumber \\
& &  + \sum_{q=1}^\infty c_{2q}^\rho \Delta_{xy} \sin(
2q\vartheta_{xy} \!+\!
qQxy) \bigg] \nonumber \\
& & \!\!\!\!\!\!\!+ h^\varepsilon_{xy}\big[
c_0^\varepsilon\dot{\vartheta}_{xy}^2 + \sum_{q=1}^\infty
c_{2q}^\varepsilon \cos( 2q\vartheta_{xy}+qQxy)\big]\bigg\},
\end{eqnarray}
where $c_{2q}^{\rho/\epsilon}$ are constants determined by the
high-energy physics. Here we have kept only
the single-plaquette terms in the charge density, and the
single-site terms in the energy density. Note that the operators
multiplying $h^\rho$ involve sines rather than cosines, which is a
consequence of particle-hole and reflection symmetries.  The {\sl
form} in Eq.~(\ref{eq:sh}) can be derived perturbatively in $v_{2q}$
using an explicit one-step coarse-graining procedure\cite{unpub}.
The quantitative reliability of this constructive
procedure is, however, limited to small bare $v_{2q}$, so for the
problem of interest the constants $c_{2q}^{\rho/\varepsilon}$
should be viewed as phenomenological parameters. Eq.~(\ref{eq:sh})
should be interpreted analogously to the non-trivial bosonization
rules for e.g. the charge density and other operators in one
dimension. Very generally, a low energy effective action must be
accompanied by rules specifying the {\sl operator content} of
various physical fields.  In analogy to the notations used there,
one may rewrite Eq.~(\ref{eq:sh}) as a scaling equality between the
microscopic and coarse-grained fields, as in Eq.~(\ref{eq:opcontentn}).

\subsection{Self-dual form - Bosonization analogy}

Having established the form of the effective action for the
$\vartheta$ variables, we are also interested in being able to compute
quantities involving the boson phase $\varphi$.  Since the effective
theory is perturbative in ${\cal S}_1$, it is straightforward to
re-introduce $\varphi$ using {\sl Gaussian} integration.  In
particular, the quadratic action ${\cal S}_0$ in Eq.~(\ref{Gaussian})
with the kernel in Eq.~(\ref{kernel}) can be transformed into
the form shown in Eq.~(\ref{eq:selfdual}) in the introduction using a
Hubbard-Stratonovich transformation:
\begin{eqnarray}\label{eq:hsvarphi}
  e^{-\frac{1}{2}\int_{{\bf k}\tau}  \frac{1}{ \pi^2 {\cal K}({\bf k}) }
   | \partial_\tau \vartheta({\bf
      k},\tau) |^2} & = & \nonumber \\
  &&\hspace{-2.0in} \int\![d\varphi] e^{-\frac{1}{2} \int_{{\bf k}\tau}  {\cal K}({\bf k}) |(\Delta_{xy} \varphi)_{\bf k}
     |^2}e^{\int_\tau \sum_{\bf r} \frac{i}{\pi} \partial_\tau \vartheta_{\bf
      r'} \Delta_{xy}\varphi_{\bf r}},
\end{eqnarray}
where ${\bf r}^\prime = {\bf r} + \hat{x}/2 + \hat{y}/2$ 
as usual, and $(\Delta_{xy}
\varphi)_{\bf k}$ denotes the Fourier transform of $\Delta_{xy}
\varphi_{\bf r}$.  After an integration by parts in the last term on
the right hand side above, the full Gaussian part of the action takes
a particularly transparent form, $S_0 = \int_\tau {\cal L}_0$, with
Lagrangian ${\cal L}_0$ defined in Eqs.~(\ref{eq:selfdual},\ref{eq:effham}).

The partition function is now represented as a path integral over
both sets of ``low energy'' fields, $\varphi_{\bf r}$ and $\vartheta_{\bf r}$,
in an appealing self-dual form.  The Gaussian theory above
gives a fixed point description of the Bose metal phase.
When augmented by the operator content of the fields, as in Eq.~(\ref{eq:opcontentn}),
together with the irrelevant non-linear interaction terms
in Eqs.~(\ref{eq:onsite},\ref{eq:twosite}),
it gives
a complete description of the universal properties of the Bose metal phase,
as detailed in the next Section. 

It will sometimes be convenient to integrate out the dual field, $\vartheta$,
leaving a Gaussian action just in terms of the phase of the boson
wavefunction:
\begin{equation}
\label{Gaussianphi}
  {\cal S}_0 = {1 \over 2} \int_{\bf k}\int_\omega{\cal M}_\varphi({\bf k},
  \omega)  | \varphi({\bf k},\omega) |^2,
\end{equation}
with Kernel,
\begin{equation}
\label{kernelphi}
  {\cal M}_\varphi({\bf k},\omega) = \frac{\omega^2 + E_{\bf k}^2}{{\cal U}({\bf k})},
\end{equation}
and with $E_{\bf k}$ as given in Eq.~(\ref{modeenergy}).
Notice that this form is identical to the Gaussian fixed point theory in terms of $\vartheta$ in Eq.~(\ref{Gaussian}), except that
${\cal U}({\bf k})$ has replaced $\pi^2 {\cal K}({\bf k})$ in
the denominator of Eq.~(\ref{kernelphi}).

The above self-dual representation makes particularly apparent
the close analogy between this theory and Bosonization theory
in one-dimension\cite{Lutt3}.
In particular, underlying both theories is a pair of dual scaler fields,
one the phase of the wavefunction and the other related to the density.
The Luttinger liquid fixed point is a Gaussian theory in the two fields,
and has a form which is nearly identical to the Bose metal action above:
\begin{equation}
{\cal L}_{1d} = {\cal H}_{1d} + \frac{i}{\pi} \int_x \partial_\tau \phi
\partial_x \theta ,
\end{equation}
with Gaussian Hamiltonian:
\begin{equation}
{\cal H}_{1d} = \int_x [ \frac{K}{2} (\partial_x \phi)^2 + \frac{U}{2\pi^2}
(\partial_x \theta)^2  ]   .
\end{equation}
Notice that the long-wavelength particle density in Bosonization theory, $n \sim \partial_x \theta/\pi$, has simply been replaced by a (lattice)
second derivative in our $2+1$-dimensional theory:  $n \sim \Delta_{xy}
\vartheta/\pi$.  The commutation relations between the two dual fields has
also been modified, as is apparent by inspecting the
Berry's phase term involving both fields in the above
Lagrangians.
The expressions in Eq.~(\ref{eq:opcontentn}) relating the bare boson density
and energy density to the low-energy fields are a generalization
of the more familiar Bosonization expressions, where for example the
Boson density near $2k_F$ is proportional to $\cos(2\theta)$.

A key strength of Bosonization in 1d is that it allows
one to study the instabilities of the Luttinger liquid towards
various types of ordered phases, such as a charge density wave state.
Similarly, the above Gaussian representation of the Bose metal
fixed point is particularly suitable for studying instabilities
both towards insulating states with broken translational
symmetry and towards a superfluid.
But before undertaking this analysis, we explore the
Bose metal phase in some detail.

\section{The Bose Metal phase\label{sec:bosemetal}}

\renewcommand{\ni}{\noindent}

We now turn to the physical properties of the Bose metal phase, using
the effective low energy theory developed in the previous section.
For the present, we will assume stability of the Bose metal, so that
all physical quantities are accessible via 
perturbation theory (in ${\cal S}_1$) around the Gaussian theory for
$\varphi,\vartheta$.  The validity of this assumption is discussed in
Sec.~\ref{sec:instab}.   

\subsection{Thermodynamic properties}

Consider first the thermodynamic properties of the Bose metal.  The
simplest is the compressibility $\kappa=\partial n/\partial \mu$.
This is trivially given from the Gaussian theory, and one finds
$\kappa=1/{\cal U}({\bf k}={\bf 0})$.  Numerically, the compressibility may be
obtained from an extrapolation of the density-density correlation
function,
\begin{equation}
  \chi_{nn}({\bf k},\omega_n) = \sum_{\bf r}\int\! d\tau\,
  \frac{1}{\pi^2}\langle \Delta_{xy}\theta_{\bf r}(\tau)
  \Delta_{xy}\theta_{\bf 0}(0)\rangle e^{i\omega_n\tau - i {\bf
      k}\cdot{\bf x}},
\end{equation}
from which $\kappa=\chi_{nn}({\bf k}\to 0,\omega=0)$.
The compressibility $\kappa$ determined in this way from the
simulations were used to define the phase diagram in
Fig.~(\ref{fig:phasediag}) 

Also interesting is the specific heat.  Since the Bose metal is
essentially a free boson theory, this is determined completely from
the density of states for the {\sl collective} free boson modes
\begin{equation}
  \rho(E) = \int_{\bf k} \delta(E-E_{\bf k}),
\end{equation}
where the boson energy is
\begin{equation}
  E_{\bf k}= \Omega({\bf k}) 4|\sin \frac{k_x}{2} \sin \frac{k_y}{2}|,
\end{equation}
and $\Omega({\bf k}) = \sqrt{{\cal U}({\bf k}) {\cal K}({\bf k})}$.
Like an ordinary (Fermi liquid) metal, the density of states gets a
large finite contribution from the low energy modes near the ``Bose
surface'' along the coordinate axes in momentum space.  Unlike a Fermi
liquid, however, there is a weak logarithmic divergence associated
with the crossing point at $k_x=k_y=0$.  In particular, 
\begin{equation}
  \rho(E) \sim \frac{1}{\pi^2\Omega_0} \left[ \ln (1/\epsilon) +
    {\cal C}_0 + O(\epsilon)\right],
\end{equation}
where $\Omega_0=\Omega(0,0)$, $\epsilon = E/\Omega_0$ and
\begin{equation}
  {\cal C}_0 = 4\ln 2 + \int_0^\pi\! \frac{dk}{\sin k/2}
  \left(\frac{\Omega_0}{\Omega(k,0)} - 1\right).
\end{equation}
Note that the constant term in the density of states depends upon the
full form of the dispersion $\Omega(k,0)=\Omega(0,k)$ all along the
Bose surface, while the logarithmic term depends only upon the
behavior at ${\bf k=0}$.  The specific heat is then
\begin{equation}
C_v = \frac{T}{4} \int_0^\infty \! dx\, \frac{x^2\rho(Tx)}{\sinh^2 x/2},
\end{equation}
which to the same accuracy gives
\begin{equation}
C_v \sim \frac{T}{4\pi^2\Omega_0}\left[ I_0 \ln
  \frac{\Omega_0}{T} + {\cal C}_0 I_0+I_1\right],
\end{equation}
and $I_n = \int_0^\infty \! dx\, x^2 (\ln(1/x))^n/\sinh^2(x/2)$, or
$I_0=4\pi^2/3$, $I_1=2 (3-2\gamma_E)\pi^2/3 + 8 \zeta'(2) \approx
4.643$, where $\zeta(z)$ is the Riemann zeta function, and
$\gamma_E \approx 0.5772$ is Euler's constant.  


\subsection{Boson correlation functions} 

It is interesting to contrast the large density of states for
collective excitations with the tunneling density of states for the
original bosons, which we will see is strongly suppressed.  In
particular, consider
\begin{equation}
  G_\phi({\bf r},\tau) = \langle e^{i\phi_{\bf
      r}(\tau)} e^{-i\phi_{\bf 0}(0)} \rangle.
\end{equation}
By symmetry, since the boson number is conserved on each row and
column, $G_\phi({\bf r},\tau)$ can be non-zero only for ${\bf r=0}$.
To further determine the behavior of $G_\phi$ in the Bose metal, we
must relate the microscopic Bose creation/annihilation operators to
the low energy modes.  By symmetry,
\begin{equation}
  e^{i\phi_{\bf r}} \sim e^{i\varphi_{\bf
      r}}\left[1+A\cos(2\vartheta_{\bf r}+2\pi\bar{n}xy)+\cdots\right].
\end{equation}
We expect the low energy properties to be dominated by the first term,
i.e. simply replacing $\phi\rightarrow\varphi$.  Since $\varphi$ is a
Gaussian variable, 
\begin{equation}
  G_\phi({\bf r},\tau) \sim e^{-F^\phi(\tau)} \delta_{\bf r,0},
\end{equation}
with 
\begin{equation}
F^\phi(\tau) = \frac{1}{2}\langle(\varphi_{\bf 0}(\tau)-\varphi_{\bf
  0}(0))^2\rangle=\int_{\bf k}\frac{{\cal U}({\bf k})}{2E_{\bf
    k}}\left(1-e^{-E_{\bf k}\tau}\right). \label{eq:Fdef}
\end{equation}
Since $E_{\bf k}$ vanishes linearly {\sl both} as $k_x\rightarrow 0$
{\sl and} as $k_y\rightarrow 0$, inspection of Eq.~(\ref{eq:Fdef})
shows that $F^\phi(\tau)$ has a {\sl double} logarithmic divergence as
$\tau\rightarrow \infty$.  Indeed, a careful calculation shows that 
\begin{equation}
  F^\phi(\tau) \sim \frac{F_2^\phi}{2} |\ln \Omega_0\tau|^2 + F^\phi_1 \ln
  (\Omega_0\tau) +\cdots
\end{equation}
for $\Omega_0\tau \gg 1$, with
\begin{eqnarray}
  \label{eq:tdoscoeffs}
  F_2^\phi & = & \frac{1}{2\pi^2}\sqrt{\frac{{\cal U}_0}{{\cal
        K}_0}} , \\
  F_1^\phi & = & \frac{1}{\pi^2} \int_0^\pi\! dk\,
  \left[\sqrt{\frac{{\cal U}(k,0)}{{\cal K}(k,0)}} \frac{1}{2\sin
      k/2} - \sqrt{\frac{{\cal U}_0}{{\cal
          K}_0}}\frac{1}{k}\right] \nonumber \\
  & & - \frac{1}{2\pi^2}\sqrt{\frac{{\cal U}_0}{{\cal
        K}_0}} \left[ 2\ln \pi +\gamma_E\right],
\end{eqnarray}
where ${\cal U}_0={\cal U}(0,0)$, ${\cal K}_0={\cal K}(0,0)$.  The $\ln^2 \Omega_0\tau$
behavior of $F^\phi(\tau)$ at large $\tau$ implies that $G_\phi(\tau)$
decays faster than any power law.  This translates into a similar
singular behavior for the tunneling density of states.  Writing the
boson Green's function in a spectral representation,
\begin{equation}
  G_\phi({\bf 0},\tau) = \int_0^\infty \! dE\, \rho_{\rm tun}(E) e^{-E|\tau|},
\end{equation}
one finds using a simple saddle-point analysis that the above behavior
at large $\tau$ requires
\begin{equation}
  \rho_{\rm tun}(E) \sim \exp\left\{-\frac{\alpha}{2}\ln^2
    \frac{\Omega_0}{E} - \ln \frac{\Omega_0}{E} (\beta\ln\ln
    \frac{\Omega_0}{E} + \lambda) \right\},
\end{equation}
where $\alpha =\beta=F_2^\phi$, $\lambda=F_1^\phi-1-F_2^\phi+\ln F_2^\phi$.

Thus, although the Bose metal possesses a large
set of gapless modes, the density of states for adding a boson into
the system vanishes at low energy. Again, this is analogous to a
Luttinger liquid in 1D\cite{Lutt3}; the conservation of particles per row or
column means that the added particle affects all the particles in a
particular row or column, leading to a suppressed amplitude for such
tunneling events.  Remarkably, the tunneling density of states
actually vanishes {\sl faster} than in a Luttinger liquid, indeed
faster than any power law in energy.  This behavior can be roughly
understood as arising from two orthogonality catastrophes occurring
simultaneously in the row and column in which the boson is added or
removed.  

It is also instructive to consider the
boson four-point correlation function, for simplicity at equal times.
Due to the row/column symmetries this is non-vanishing
only when the four points sit at the corners of a
rectangle:
\begin{eqnarray}
\!\!G_\phi^{(4)}(x,y)&=&\!\!\left\langle e^{i\phi_{oo}(\tau)}
  e^{i\phi_{xo}(\tau)} e^{i\phi_{oy}(\tau)}
  e^{i\phi_{xy}(\tau)} \right\rangle.  
\end{eqnarray}
Upon replacing $\phi$ with the low energy field $\varphi$,
this can be readily evaluated using the Gaussian Bose metal action.
For any {\sl fixed} $y$, one finds power-law behavior in $x$ (and vice versa since
$G_\phi^{(4)}(y,x)=G_\phi^{(4)}(x,y)$) 
\begin{equation}
  G_\phi^{(4)}(x,y) \sim\frac{1}{|x|^{\eta(y)}}\qquad
  {\rm for}\, |x| \gg 1, \label{eq:gphigen}
\end{equation}
with
\begin{equation}
  \eta(y) = \frac{1}{\pi^2} \int_0^\pi\! dk\, \sqrt{\frac{{\cal U}(0,k)}{{\cal K}(0,k)}}
  \frac{\sin^2 (ky/2)}{\sin (k/2)}.
\label{etaofy}
\end{equation}
Note that
Eq.~(\ref{eq:gphigen}) requires only $|x| \gg 1$ and places no
restriction on $y$.  Hence it obtains also when {\sl both} arguments
are large, and thus
\begin{equation}
  G_\phi^{(4)}(x,y) \sim  e^{- \frac{1}{\pi^2} \sqrt{\frac{{\cal
          U}_0}{{\cal K}_0}} (\ln x)(\ln y) - C (\ln x + \ln y)},
\end{equation}
as $x,y \rightarrow \infty$, with
\begin{eqnarray}
  C & = & \frac{1}{\pi^2}\sqrt{\frac{{\cal U}_0}{{\cal
        K}_0}}(\gamma_E+\ln\pi) \nonumber \\
  & & + \frac{1}{\pi^2}\int_0^\pi \!
  \left(\sqrt{\frac{{\cal U}(0,k)}{{\cal K}(0,k)}} \frac{1}{2\sin k/2} 
    - \sqrt{\frac{{\cal U}_0}{{\cal
        K}_0}}\frac{1}{k}\right).
\end{eqnarray}

\subsection{Vortex correlation functions}
The exact plaquette duality makes it possible to define a number of
characteristic ``vortex'' correlators in terms of the field
$e^{i\theta}$.  As for the boson correlator above, the behavior in the
Bose metal depends upon the expression for the vortex operator in the
low energy variables.  By symmetry, we expect the leading terms
(involving the smallest exponentials of $\vartheta$) to be
\begin{equation}
  e^{i\theta_{\bf r}} \sim {\rm Re}\left[ A e^{i\pi\bar{n}xy}
    e^{i\vartheta_{\bf r}} \right] , \label{eq:thetageneral}
\end{equation}
where in general $A$ is complex.  Note that, since $\theta/\pi$ is an
integer, the microscopic vortex field satisfies
$e^{i\theta}=e^{-i\theta}$, but this is not true for $\vartheta$.  In
the special case of $\bar{n}=1/2$, particle-hole symmetry
($\vartheta\rightarrow-\vartheta$) further
implies $A$ is real, whence
\begin{equation}
  e^{i\theta_{\bf r}} \sim_{\bar{n}=1/2} 2A \cos(\vartheta_{\bf
    r}-\frac{\pi}{2}xy). \label{eq:thetaph}
\end{equation}

\begin{figure}
\begin{center}
\vskip-2mm
\hspace*{0mm}
\centerline{\fig{2.8in}{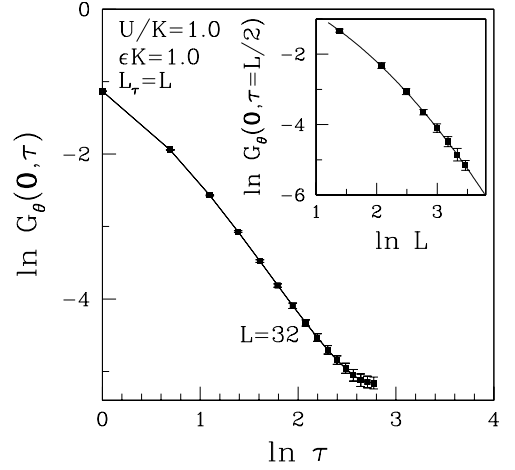}}
\vskip-2mm
\caption{Vortex correlation function $G_\theta({\bf 0},\tau)
= \left\langle e^{i\theta_{\bf r}(\tau)} e^{-i\theta_{\bf 0}(0)}
\right\rangle$ in the Bose metal phase.  The
correlation function decays faster than a power law $\sim \exp \left(-
\frac{1}{4}\sqrt\frac{{\cal K}_0}{{\cal U}_0} |\ln \Omega_0 \tau|^2
\right)$ at
long times (see text). Inset shows this decay in a finite size scaling
plot ($L=4$ to $L=32$) of $\ln G_\theta({\bf 0},\tau)$ versus $\ln
\tau$ for $\tau = L_{\tau}/2$, from which we extract $\frac{{\cal U}_0}{{\cal K}_0} 
\approx 1$, nearly its bare value for these parameters.}
\label{fig:bmettau}
\end{center}
\end{figure}

With this relation in hand, let us consider first the vortex two-point
function,
\begin{equation}
  G_\theta({\bf r},\tau) = \left\langle e^{i\theta_{\bf r}(\tau)}
  e^{-i\theta_{\bf 0}(0)}\right\rangle.
\end{equation}
As for the boson correlator $G_\phi$, the vortex two-point functions
vanishes unless ${\bf r=0}$ due to the dual row and column symmetries.
 Evaluating it using Eq.~(\ref{eq:thetageneral}) and the Gaussian
action, one finds similar results to the boson tunneling density of
states, 
\begin{equation}
\label{eq:vortcorr}
  G_\theta({\bf 0},\tau) \sim \exp \left\{-\frac{{F}_2^\theta}{2} |\ln
    \Omega_0 \tau|^2 -{F}_1^\theta |\ln\Omega_0\tau|\right\},
\end{equation}
where 
\begin{eqnarray}
  \label{eq:thetadoscoeffs}
  F_2^\theta & = & \frac{1}{2}\sqrt{\frac{{\cal K}_0}{{\cal
        U}_0}} , \\
  F_1^\theta & = &  \int_0^\pi\! dk\,
  \left[\sqrt{\frac{{\cal K}(k,0)}{{\cal U}(k,0)}} \frac{1}{2\sin
      k/2} - \sqrt{\frac{{\cal K}_0}{{\cal
          U}_0}}\frac{1}{k}\right] \nonumber \\
  & & - \frac{1}{2}\sqrt{\frac{{\cal K}_0}{{\cal
        U}_0}} \left[ 2\ln \pi +\gamma_E\right].
\end{eqnarray}
In fact, these coefficients are obtained directly from
Eq.~(\ref{eq:tdoscoeffs}) using the duality transformation ${\cal
  K}({\bf k}) \leftrightarrow {\cal U}({\bf k})/\pi^2$.  In
Fig.~\ref{fig:bmettau} we show numerical results for the two-point
vortex correlation function obtained from the quantum Monte Carlo
simulation in the parameter regime corresponding to the Bose metal
phase.  The downward curvature of the data is consistent with a decay
more rapid than a power law, and as shown in the inset can be fit to
the form in Eq.~(\ref{eq:vortcorr}) with $\frac{{\cal U}_0}{{\cal K}_0}
\approx 1$.

\begin{figure}
\begin{center}
\vskip-2mm
\hspace*{0mm}
\centerline{\fig{2.8in}{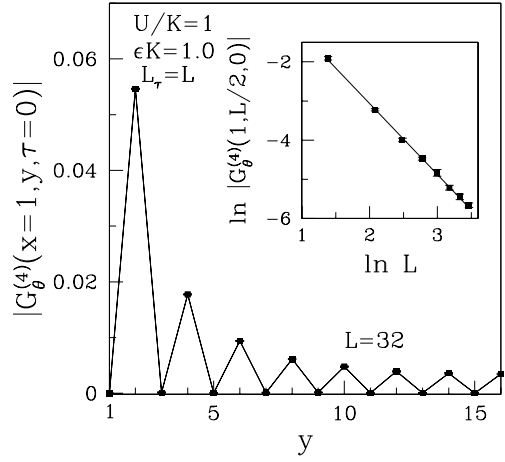}}
\vskip-2mm
\caption{The 4-point vortex correlation function 
$G^{(4)}_\theta(x=1,y,\tau=0)$, defined in the text, evaluated in
the Bose metal phase. The correlation function vanishes for
odd values of $y$ due to particle-hole symmetry at $\bar{n}=1/2$.
The expected power law decay of the envelope is shown in the
inset as a finite size scaling plot, from which we extract the
exponent ${\eta_v}(1) \approx 1.9 \pm 0.1$, close to its bare
value for $U/K=1$.}
\label{fig:bmetxy}
\end{center}
\end{figure}

Next consider the vortex four-point function,
which due to the dual row/column symmetry is non-vanishing only when the four points sit at the corners of a
rectangle:
\begin{eqnarray}
\!\!G_\theta^{(4)}(x,y)&=&\!\!\left\langle e^{i\theta_{oo}(\tau)}
  e^{i\theta_{xo}(\tau)} e^{i\theta_{oy}(\tau)}
  e^{i\theta_{xy}(\tau)} \right\rangle.  
\end{eqnarray}
This can be re-expressed in terms of the low-energy field, $\vartheta$,
using Eq.~(\ref{eq:thetageneral}), and then evaluated with the Gaussian
Bose metal action.  
For any {\sl fixed} $y$, one finds power-law behavior in $x$,
\begin{equation}
  G_\theta^{(4)}(x,y) \sim\frac{\cos (\pi\bar{n}xy)}{|x|^{{\eta_v}(y)}}\qquad
  {\rm for}\, |x| \gg 1, \label{eq:gthetagen}
\end{equation}
with
\begin{equation}
  {\eta_v}(y) = \int_0^\pi\! dk\, \sqrt{\frac{{\cal K}(0,k)}{{\cal U}(0,k)}}
  \frac{\sin^2 (ky/2)}{\sin (k/2)}.
\end{equation}
Interestingly, one can see from Eq.~(\ref{eq:gthetagen}) that for the
case $\overline{n}=1/2$, the four boson correlator {\sl vanishes}
exactly whenever $xy$ is odd as a consequence of particle-hole
symmetry.  This behavior, and the associated power-law correlations,
are shown in Fig.~\ref{fig:bmetxy}.  In the limit when both
$x,y \rightarrow \infty$, the vortex four-point correlator vanishes faster than any power law:
\begin{equation}
  G_\theta^{(4)}(x,y) \sim \cos (\pi\bar{n}xy) e^{- \sqrt{\frac{{\cal
          K}_0}{{\cal U}_0}} (\ln x)(\ln y)- C_v (\ln x + \ln y)},
  \label{eq:G4theta} 
\end{equation}
as $x,y \rightarrow \infty$, with
\begin{eqnarray}
  C_v & = &\sqrt{\frac{{\cal K}_0}{{\cal
        U}_0}}(\gamma_E+\ln\pi) \nonumber \\
  & & + \int_0^\pi \!
  \left(\sqrt{\frac{{\cal K}(0,k)}{{\cal U}(0,k)}} \frac{1}{2\sin k/2} 
    - \sqrt{\frac{{\cal K}_0}{{\cal
        U}_0}}\frac{1}{k}\right).
\end{eqnarray}
With the exception of the $\cos(\pi\bar{n}xy)$ prefactor in
Eq.~(\ref{eq:G4theta}), all the results in this subsection can be
obtain from those in the previous one by the duality transformation
$\varphi \leftrightarrow\vartheta$ and ${\cal U}({\bf k})
\leftrightarrow \pi^2 {\cal K}({\bf k})$.

\subsection{Collective correlation functions}

Next, let us consider the correlations of ``two-particle''
operators such as the boson and energy densities.
As above, we require the operator correspondences between the
microscopic and effective variables.  These were already worked out in Sec.~\ref{sec:effective},
and summarized in Eqs.~(\ref{eq:opcontentn}-\ref{eq:opcontente}) in
the introduction. For simplicity, we focus on the most interesting
case of $\overline{n}=1/2$, for which $Q=\pi$, and furthermore keep
only the lowest non-trivial harmonics with $q=1$.  In this limit,
Eqs.~(\ref{eq:opcontentn}-\ref{eq:opcontente}) reduce to
\begin{eqnarray}\label{eq:opcontentnhalf}
\delta n_{x+\frac{1}{2},y+\frac{1}{2}} & \sim & c_0^\rho
\Delta_{xy}\vartheta_{xy}
 \nonumber \\
& &   \hspace{-0.65in}+ c_{2}^\rho \sum_{a,b=0,1}
(-1)^{(x+a)(y+b)+a+b} \sin( 2\vartheta_{x+a,y+b}) , \\
\varepsilon_{xy} & \sim & c_0^\varepsilon\dot{\vartheta}_{xy}^2 +
c_{2}^\varepsilon (-1)^{xy}\cos( 2\vartheta_{xy}).
\label{eq:opcontentehalf} 
\end{eqnarray}
Consider first the density-density correlation function,
\begin{equation}
  \chi_{nn}({\bf r},\tau) = \langle
  \delta n_{x+\frac{1}{2},y+\frac{1}{2}}(\tau)
  \delta n_{\frac{1}{2},\frac{1}{2}}(0)\rangle. 
\end{equation}
Substituting for $\delta n$ using Eq.~(\ref{eq:opcontentnhalf}), one
obtains three contributions to $\chi_{nn}$:
\begin{equation}
\chi_{nn}({\bf r},\tau) \sim (c_0^\rho)^2 \chi_{nn}^{00}({\bf
  r},\tau)+ (c_2^\rho)^2 \chi_{nn}^{22}({\bf
  r},\tau)+c_0^\rho c_2^\rho \chi_{nn}^{20}({\bf
  r},\tau) .
\end{equation}
The cross term $\chi_{nn}^{20}$ is negligible.  The
first contribution, $\chi_{nn}^{00}({\bf r},\tau)$ is just the
correlator between ``coarse-grained'' densities $\propto
\Delta_{xy}\vartheta_{xy}$.  This term is non-zero in the Gaussian
theory, and gives a smooth function of ${\bf r}$, with a power-law
behavior at large arguments.  For instance, at equal times and large
$|x|,|y|\gg 1$,
\begin{equation}
  \chi_{nn}^{00}({\bf r},0) \sim \sqrt{\frac{{\cal K}_0}{{\cal
        U}_0}}\frac{1}{x^2y^2}. 
\end{equation}
More generally, $\chi_{nn}^{00}$ has a smooth Fourier transform,
\begin{equation}
  \chi_{nn}^{00}({\bf q},\omega_n) =  \frac{\pi^2}{{\cal U}({\bf k})}
  \frac{E_{\bf k}^2}{\omega_n^2 + E_{\bf k}^2}.
\end{equation}
Perturbative corrections from ${\cal S}_1$ to $\chi_{nn}^{00}$ do not
modify this qualitative behavior.  

The remaining contribution to the density-density correlator comes
from the $\sin 2\vartheta$ terms in Eq.~(\ref{eq:opcontentnhalf}).
Naively, this contribution is ultra-local (and hence uninteresting),
i.e. vanishes unless $|x|,|y| \leq 1$, as a consequence of the fact
that the discrete row/column symmetries are promoted to continuous
ones at the Gaussian level.  One may interpret this as meaning that
the ``vorticity'' on each row or column is conserved exactly in the
Gaussian model.  This conclusion, however, is incorrect
once the non-quadratic corrections in ${\cal S}_1$ are taken into
account, since expanding factors of $\upsilon_2
(-1)^{xy}\cos2\vartheta$ can ``supply'' vorticity in units of $2$ to a 
particular site.  Hence, $\chi_{nn}^{22}$ becomes non-trivial at
second order in $\upsilon_2$.  Provided $|x|,|y| >1$, the appropriate
second order perturbative expression is 
\begin{eqnarray}
  \chi_{nn}^{22}({\bf r},\tau) &  =  -\frac{\upsilon_2^2}{8}
  \sum_{abcd=0,1} & (-1)^{(x+c-a)(y+d-b)+a+b+c+d}  \nonumber \\
&& \times C^{(4)}_{x+c-a,y+d-b}(\tau), \label{eq:ptorder2}
\end{eqnarray}
where
\begin{equation}
  C^{(4)}_{x,y}(\tau) = \int\!d\tau_1d\tau_2\, \left\langle 
    e^{2i[\vartheta_{oo}(0)+\vartheta_{xy}(\tau)-\vartheta_{x0}(\tau_1)-
      \vartheta_{0y}(\tau_2)]} \right\rangle_0,     
\end{equation}
where the expectation value indicates an average calculated in the
Gaussian theory.  At large distances, $|x|,|y| \gg 1$, one expects
$C^{(4)}_{x+c-a,y+d-b}(\tau) \approx C^{(4)}_{x,y}(\tau)$, independent of
$a,b,c,d$.  In this limit, only the prefactor in
Eq.~(\ref{eq:ptorder2}) depends upon $a,b,c,d$, and the sums can be
carried out explicitly.  To do so, it is convenient to employ the
identity 
\begin{equation}
\label{intidentity}
  (-1)^{xy} = \frac{1}{2}[1+(-1)^x+(-1)^y - (-1)^{x+y}],
\end{equation}
valid for integer $x,y$.  Applying this identity to
Eq.~(\ref{eq:ptorder2}), only the last term survives the sum, and
gives
\begin{equation}
  \chi_{nn}^{22}({\bf r},\tau) \approx \upsilon_2^2 (-1)^{x+y}
  C^{(4)}_{x,y}(\tau), 
\end{equation}
for $|x|,|y| \gg 1$.  This indicates the presence of $(\pi,\pi)$
correlations in the boson density.  
More generally, if only $|x| \gg 1$ but $|y|>1$ but still of $O(1)$,
one finds
\begin{equation}
\chi_{nn}^{22}({\bf r},\tau) \approx \frac{\upsilon_2^2}{4} (-1)^{x+y}
\sum_{bd=0,1} (1-(-1)^{y+d-b}) C^{(4)}_{x,y+d-b}(\tau), \label{eq:chinnfixedy}
\end{equation}
so at {\sl any} fixed $y$ the density-density correlator oscillates at 
wavevector $\pi$ as a function of $x$ (and vice-versa by rotational
symmetry).  To establish the range of these
correlations, we must consider $C^{(4)}_{x,y}(\tau)$ in some detail.
Using properties of Gaussian fields, we have
\begin{equation}
  C^{(4)}_{x,y}(\tau) = \int\!d\tau_1 d\tau_2\, \exp\left[ -
  c_{x,y}(\tau,\tau_1,\tau_2) \right], \label{eq:bigCxy}
\end{equation}
where
\begin{eqnarray}
  c_{x,y}(\tau,\tau_1,\tau_2) & = &  \\
  & & \hspace{-0.6in} 2\left\langle \left(
      \vartheta_{xy}(\tau)\!+\!\vartheta_{00}(0)\!-\!
      \vartheta_{x0}(\tau_1)\!-\!
      \vartheta_{0y}(\tau_2)\right)^2\right\rangle_0 . \nonumber
\end{eqnarray}

The calculation and analysis of $c_{xy}$ is somewhat involved, and is
described in detail in Appendix~\ref{sec:asymptotics}.
There we derive a useful
approximation for $c_{xy}$ which captures the behavior in all the
relevant limits ($|x|\gg 1,\Omega_0|\tau-\tau_1|,\Omega_0|\tau_2|$, and $|y|<|x|$):
\begin{eqnarray}
  c_{xy} & \sim & 2{\eta_v}(y)\ln(x^2\!+\!\Omega_y^2 \tau^2) \nonumber 
  \\
  & & +  \sqrt{\frac{{\cal
    K}_0}{{\cal U}_0}} \left( {\cal
    F}[\frac{\Omega_0|\tau\!-\!\tau_1|}{|y|}]+{\cal
    F}[\frac{\Omega_0|\tau_2|}{|y|}] \right), \label{eq:cxyasympt}
\end{eqnarray}
where $\Omega_y \sim \Omega_0$ (but in general depends weakly upon the 
full form of ${\cal K}(0,k_y),{\cal U}(0,k_y),$ and $y$) and the
crossover function ${\cal F}({\cal X})$ satisfies
\begin{equation}
  {\cal F}({\cal X}) \approx \cases{0 & ${\cal X} \ll 1$ \cr 
    \ln^2{\cal X} & ${\cal X} \gg 1$}. \label{eq:Fcross}
\end{equation}
Using Eqs.~(\ref{eq:cxyasympt},\ref{eq:Fcross},\ref{eq:bigCxy}), we
find
\begin{equation}
  C^{(4)}_{xy}(\tau) \stackrel{\sim}{\scriptscriptstyle |x|\gg 1} \left(\frac{y}{\Omega_0}\right)^2 
  e^{2\left(\frac{{\cal K}_0}{{\cal U}_0}\right)^{1/4}} 
  \left(x^2+\Omega_y^2\tau^2\right)^{-2{\eta_v}(y)}, \label{eq:c4asymptotics}
\end{equation}
and of course the same behavior with $x\leftrightarrow y$. 

This power-law behavior of $C^{(4)}_{xy}$, and hence
$\chi_{nn}({\bf r},\tau)$ translates into singularities in
the static structure factor,
\begin{eqnarray}
  S_{nn}({\bf k}) & \equiv & \sum_{xy}e^{-i(k_x x+k_y y)}
  \chi_{nn}({\bf r},\tau=0),
\end{eqnarray} 
for wavevectors near ${\bm\pi}\equiv (\pi,\pi)$.  In particular, let ${\bf
  k}={\bm\pi}+{\bf q}$, and consider the limit $q_x \ll 1$ with $q_y$
fixed.  In this limit the {\sl singular behavior} of the structure
factor is dominated by large $|x|$ but $y$ of $O(1)$.  Hence we may
apply Eq.~(\ref{eq:chinnfixedy}) to write
\begin{eqnarray}
  S_{nn}({\bm \pi}+{\bf q}) & \stackrel{\sim}{\scriptscriptstyle |q_x|
    \ll 1} & \\  
  && \hspace{-1.0in} \upsilon_2^2 \sum_{{\rm
      odd}\, y} \cos q_y y (1+\cos q_y)\sum_x  e^{iq_x x}
  C^{(4)}_{xy}(0). \nonumber 
\end{eqnarray}
$ $From the power-law behavior in
Eq.~(\ref{eq:c4asymptotics}), one can readily see that the Fourier
transform in $x$ leads to singularities for small $q_x$.  Indeed, if
any ${\eta_v}(y)<1/4$ for any odd $y$, $\chi_{nn}$ {\sl diverges} as 
$q_x \rightarrow 0$ at fixed $q_y$, while for $1/4 < {\eta_v} < 3/4$,
the structure factor remains finite but has a divergent second
derivative at $q_x=0$:
\begin{eqnarray}
  \!\!\!S_{nn}({\bm \pi}\!+\!{\bf q}) & \!\!\sim & \!\!S_{nn}^{0}({\bf q}) \!+\!\!
  \sum_{{\rm odd}\, y} \!\!A^\rho_y(q_y) {\rm 
    sgn}(\alpha_y) |q_x|^{-\alpha_y},
\end{eqnarray}
where $S_{nn}^0({\bf q})$ is a smooth function,
$\alpha_y=1-4{\eta_v}(y)$, and $A^\rho_y(q_y)\propto \upsilon_2^2 \cos q_y y
(1+\cos q_y)$ is a
positive amplitude peaked at $q_y=0$. Hence the behavior for small $q_x$ is dominated by
the {\sl minimum} (over odd $y$) value of ${\eta_v}(y)$ (maximum of
$\alpha_y$).

\begin{figure}
\begin{center}
\vskip-2mm
\hspace*{0mm}
\centerline{\fig{2.8in}{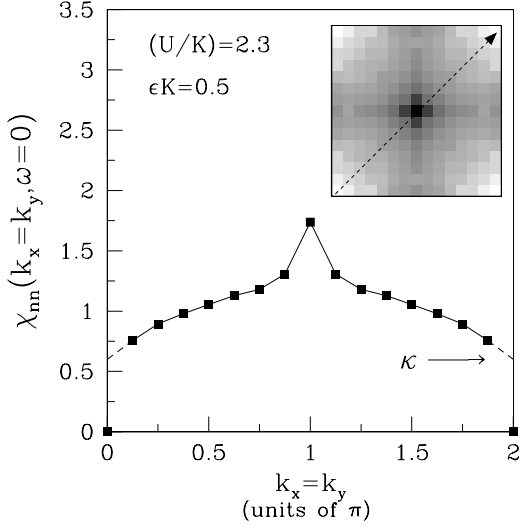}}
\vskip-2mm
\caption{The density susceptibility $\chi_{nn}({\bf
    k},\omega=0)$ in the Bose metal phase, close to the phase
  boundary, along $(0,0) \to (2\pi,2\pi)$. The cusp at $(\pi,\pi)$ is
  due to subdominant power law correlations in the Bose metal phase
  which may be viewed as arising from corrections to the density
  operator in the Gaussian theory as discussed in the paper. Inset
  shows the grayscale plot of the susceptibility over the entire
  Brillouin zone, centered on $(\pi,\pi)$ (dark regions indicate large
  susceptibility, and the $k_x=0$ and $k_y=0$ lines are omitted from
  the grayscale plot for clarity). The singular cusp is visible along
  the indicated $(0,0)\to(2\pi,2\pi)$ direction.}
\label{fig:sqrho}
\end{center}
\end{figure}

The zero frequency susceptibility,
${\chi}_{nn}({\bf k})\equiv \chi_{nn}({\bf k},\omega_n=0)$ has a similar but
stronger divergence due to the extra time integration:
\begin{eqnarray}
  \!\!\!{\chi}_{nn}({\bm \pi}\!+\!{\bf q}) & \!\!\sim &
  \!\!{\chi}_{nn}^{0}({\bf q}) \!+\!\! 
  \sum_{{\rm odd}\, y} \!\!\tilde{A}^\rho_y(q_y) {\rm 
    sgn}(\tilde{\alpha}_y) |q_x|^{-\tilde{\alpha}_y},
\label{eq:denden}
\end{eqnarray}  
where ${\chi}_{nn}^0({\bf q})$ is another smooth function,
$\tilde{A}^\rho_y(q_y)\propto A^\rho_y(q_y)$, and
$\tilde{\alpha}_y=2-4{\eta_v}(y)=1+\alpha_y$ signals the stronger
divergence.  The difference in exponents implies that if
$S_{nn}({\bm \pi+\bf q})$ has a divergent slope at $q_x=0$,
${\chi}_{nn}({\bm \pi+ \bf q})$ itself divergences there.  Conversely,
if $\tilde{\chi}_{nn}({\bm\pi+\bf q})$ has a slope divergence at $q_x=0$ (occurs for
$1/2<{\eta_v}<3/4$), the static structure factor does not.
Numerical results for the density susceptibility in the Bose metal phase
are shown in Fig.~\ref{fig:sqrho}, and reveal a singular cusp-like behavior
at wavevector ${\bm\pi}$.  This form is consistent with that predicted
by Eq.~(\ref{eq:denden}) with a maximum value $-1 < \tilde{\alpha}_y^{max} <0$.

\begin{figure}
\begin{center}
\vskip-2mm
\hspace*{0mm}
\centerline{\fig{2.8in}{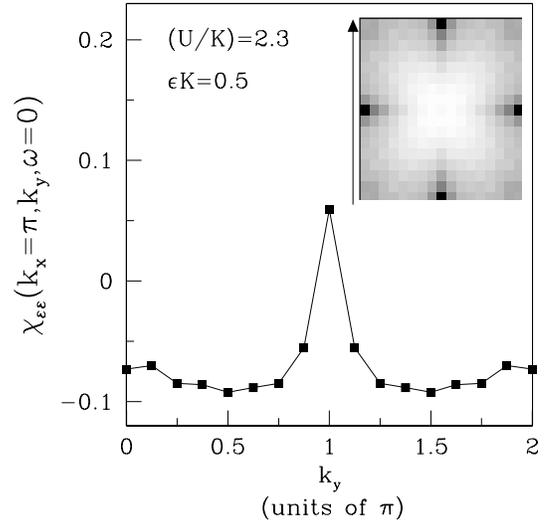}}
\vskip-2mm
\caption{The energy susceptibility $\chi_{\varepsilon\varepsilon}({\bf
    k},\omega=0)$ in the Bose metal phase, close to the phase
  boundary, along $(0,0) \to (0,2\pi)$. The cusp at $(0,\pi)$ is due
  to subdominant power law correlations in the Bose metal phase which
  may be viewed as arising from corrections to the energy-density
  operator in the Gaussian theory as discussed in the paper. Inset
  shows the grayscale plot of the susceptibility over the entire
  Brillouin zone, centered on $(\pi,\pi)$ (dark regions indicating
  large susceptibility). The singular features are most clearly
  visible as peaks (dark spots) at the $(0,\pi)$ and $(\pi,0)$ points
  and as a dip (white spot) near $(\pi,\pi)$.}
\label{fig:sqenergy}
\end{center}
\end{figure}

Now consider the energy-energy correlation function, 
\begin{equation}
  \chi_{\varepsilon\varepsilon}({\bf r},\tau) = \langle
  \varepsilon_{xy}(\tau) \varepsilon_{xy}(0) \rangle_c,
\end{equation}
where the $c$ subscript indicates the cumulant expectation value.
As for the density-density correlator, we can employ Eq.~(\ref{eq:opcontentehalf}) to expand $\chi_{\varepsilon\varepsilon}$ 
as
\begin{equation}
\chi_{\varepsilon\varepsilon}({\bf r},\tau) \sim (c_0^\varepsilon)^2
\chi_{\varepsilon\varepsilon}^{00}({\bf 
  r},\tau)+ (c_2^\varepsilon)^2 \chi_{\varepsilon\varepsilon}^{22}({\bf
  r},\tau)+c_0^\varepsilon c_2^\varepsilon \chi_{\varepsilon\varepsilon}^{20}({\bf
  r},\tau).
\end{equation}
As for the density case, $\chi^{00}_{\varepsilon\varepsilon}$ has a smooth
Fourier transform, and $\chi_{\varepsilon\varepsilon}^{20}$ is
negligible.  We focus on $\chi^{22}_{\varepsilon\varepsilon}$, which
(for $|x|,|y|>0$) to second order in $\upsilon_2$ is
\begin{equation}
  \chi_{\varepsilon\varepsilon}^{22} = \frac{\upsilon_2^2}{8}
  (-1)^{xy} C^{(4)}_{xy}(\tau).
\end{equation}
Using Eq.~(\ref{intidentity},\ref{eq:c4asymptotics}), one straightforwardly sees that
there are singularities in $\chi_{\varepsilon\varepsilon}^{22}({\bf
  k})$ as $k_x (k_y) \rightarrow 0$ and $k_x (k_y) \rightarrow \pi$.
In particular,  
\begin{eqnarray}
 \!\!\!\!\!\!S_{\varepsilon\varepsilon}({\bf k}) &
 \!\!\!\stackrel{\sim}{\scriptscriptstyle |k_x| \ll 1} & 
  \!\!\!S^0_{\varepsilon\varepsilon}({\bf k}) \!+ \!\!\!\!\sum_{{\rm
      even}\,y} \!\! A_y^\varepsilon(k_y) \,{\rm sgn}(\alpha_y)
  |k_x|^{-\alpha_y}, \\
 \!\!\!\!\!\!{\chi}_{\varepsilon\varepsilon}({\bf k}) & \!\!\!\stackrel{\sim}{\scriptscriptstyle |k_x| \ll 1} & 
  \!\!\!{\chi}^0_{\varepsilon\varepsilon}({\bf k}) \!+ \!\!\!\!\sum_{{\rm
      even}\,y} \!\! \tilde{A}_y^\varepsilon(k_y) \,{\rm
    sgn}(\tilde{\alpha}_y) |k_x|^{-\tilde{\alpha}_y} ,
\end{eqnarray}
and
\begin{eqnarray}
\label{eq:enenpi0}
 \!\!\!\!\!\!S_{\varepsilon\varepsilon}(\pi+q_x,k_y) &
 \stackrel{\sim}{\scriptscriptstyle |q_x| \ll 1} & \\
 && \hspace{-1.0in} S^0_{\varepsilon\varepsilon}(\pi+q_x,k_y) \!+ \!\!\sum_{{\rm
      odd}\,y} \!\! A_y^\varepsilon(k_y) \,{\rm sgn}(\alpha_y)
  |q_x|^{-\alpha_y}, \nonumber \\
 \!\!\!\!\!\!{\chi}_{\varepsilon\varepsilon}(\pi+q_x,k_y) &
 \stackrel{\sim}{\scriptscriptstyle |q_x| \ll 1} &  \\
&& \hspace{-1.0in}
\!\!\!{\chi}^0_{\varepsilon\varepsilon}(\pi+q_x,k_y) \!+
\!\!\sum_{{\rm odd}\,y} \!\! \tilde{A}_y^\varepsilon(k_y) \,{\rm
    sgn}(\tilde{\alpha}_y) |q_x|^{-\tilde{\alpha}_y} , \nonumber
\end{eqnarray}
where $A_y^\varepsilon(k_y) \propto \tilde{A}_y^\varepsilon(k_y)
\propto \cos k_y y$.  Similar formulae for $k_y \approx 0,\pi$ are
obtained by rotation.  Note that because $A_y^\varepsilon(k_y)\propto
\cos k_y y$ is negative for $k_y \approx \pi$ and $y$ odd, the
energy-energy correlator has a singular {\sl dip} near ${\bf k}= {\bm
  \pi}$ and singular {\sl peaks} near ${\bf k}=(0,0),(\pi,0),$ and $(0,\pi)$.
Numerical results for the energy susceptibility in the Bose metal
phase are shown in Fig.~\ref{fig:sqenergy}.  Notice the singular cusp-like
peaks at wavevectors $(0,\pi)$ and $(\pi,0)$, as predicted
in Eq.~(\ref{eq:enenpi0}).

\subsection{Electrical Conductivity}

We finally consider the electrical conductivity in the Bose metal.
To obtain an expression for the current operator
requires coupling in a vector potential.  The usual minimal
coupling prescription for a lattice model of bosons
would have one replace a discrete lattice derivative as,
\begin{equation}
\phi_{{\bf r}+{\bm \mu}} - \phi_{ \bf r} \rightarrow 
\phi_{{\bf r}+{\bm \mu}} - \phi_{ \bf r} + A_{\bf r}^{\mu}  ,
\end{equation}
with ${\bm \mu} = \hat{{\bf x}} , \hat{{\bf y}}$,
and define the current operator by differentiating with respect
to the vector potential.  For the boson ring model this prescription
is ambiguous due to the second derivative form, 
and a family of gauge inequivalent forms are possible:
\begin{equation}
\Delta_{xy} \phi_{\bf r} \rightarrow \Delta_{xy} \phi_{\bf r} + \alpha
\Delta_x A_{\bf r}^y + \beta \Delta_y A_{\bf r}^x  ,
\end{equation}
where $\Delta_x f_{\bf r}  \equiv f_{{\bf r}+\hat{\bf x}} - f_{\bf r}$
denotes a discrete derivative.
Gauge invariance requires $\alpha + \beta =1$, whence this can be
re-expressed as,
\begin{equation}
\Delta_{xy} \phi_{\bf r} \rightarrow \Delta_{xy} \phi_{\bf r} + 
\Delta_x A_{\bf r}^y + (\alpha -1) \Phi   ,
\end{equation}
where $\Phi = \Delta_x A^y - \Delta_y A^x$ is the gauge invariant flux through
the plaquette.  If one derives the boson ring model by starting with a model of electrons reformulated in terms of a $Z_2$ gauge field
coupled to spinons and chargons as detailed in Appendix ~\ref{sec:Z2}, one arrives
at an appealing symmetric form for the ring term with $\alpha=\beta=1/2$.

Henceforth we focus on the zero wavevector conductivity,
$\sigma(\omega)$.  In this case, the above ambiguity is irrelevant,
since a spatially uniform vector potential does not enter for any
value of $\alpha$.  But the vector potential will still of course
enter into the boson hopping term in Eq.~(\ref{hamhop}), and upon
differentiation generates the usual current operator,
\begin{equation}
I^\mu_{\bf r} = t \sin(\phi_{{\bf r}+{\bm \mu}} - \phi_{ \bf r} ) .
\end{equation}
When we coarse grain the theory, we should write down the
current operator in terms of the slow field, $\varphi$:
\begin{equation}
I^\mu_{\bf r} = c_I t \sin(\varphi_{{\bf r}+{\bm \mu}} - \varphi_{ \bf r} ) 
+ ...   ,
\end{equation}
with $c_I$ a dimensionless constant.  The other contributions
will include terms which hop a single boson several lattice spacings
and terms which hop several bosons - generally all local terms
allowed by the symmetries.  Such terms will generically be subdominant
at low frequencies, and so we retain only the leading contribution.

With the current operator in hand it is straightforward to obtain
the conductivity from the usual Kubo expression:
\begin{equation}
\sigma_1(\omega) = {\rm Re}\, \sigma(i \omega_n \rightarrow \omega) ,
\end{equation}
with 
\begin{equation}
\sigma(i \omega_n) = \frac{-1}{\omega_n} \int d\tau
\sum_{\bf r} e^{-i \omega_n \tau} \langle I^y_{\bf r}(\tau) I^y_{\bf 0}(0) \rangle   .
\end{equation}
We have dropped the diamagnetic contribution since it will not contribute
to the real part of the conductivity.
Evaluating the correlator using the Gaussian Bose metal action gives
at zero temperature,
\begin{equation}
\langle I^y_{\bf r}(\tau) I^y_{\bf 0}(0) \rangle_0 \sim t^2
\delta_{y,0} (x^2 + \tau^2)^{-\Delta}   ,
\end{equation}
with scaling dimension $\Delta = \eta(1)/2$ where $\eta(y)$
is given explicitly in terms of the Bose liquid parameters in Eq.~\ref{etaofy}.
Performing the time integration and spatial summation
gives a power law singular contribution,
\begin{equation}
\omega_n \sigma(i\omega_n) = - A t^2 (-1)^{{\rm int}(\Delta)}
\omega_n^{2\Delta -2} + 
\Sigma_{reg}(i\omega_n)   ,
\end{equation}
where $A$ is a positive constant and $\Sigma_{reg}$ is analytic in it's
argument and thus does not contribute to the real part of the
conductivity.  Analytic continuation to real frequencies gives a
singular contribution to the complex conductivity of the form,
\begin{equation}
\sigma(\omega) = At^2 (-1)^{{\rm int}(\Delta)} (-i \omega)^{2\Delta
  -3} + i\sigma_2(\omega)   , 
\end{equation}
with real $\sigma_2(\omega)$.  
We note that causality places
strong constraints on the phase angle\cite{dorsey} from the
singular contribution, consistent with the above form.
Finally, taking the real part gives the optical conductivity,
\begin{equation}
\sigma_1(\omega) = A t^2 |\sin(\pi \Delta)| |\omega|^{2\Delta -3}  .
\end{equation}
Notice that the amplitude of this contribution vanishes 
for integer $\Delta$.  For these special cases, the higher order
contributions to the current operator should be kept, and will
contribute a similar form but generally with larger scaling dimensions.

As we discuss in the next section, stability of the Bose metal phase
to boson hopping requires that $\Delta >2$, implying an optical
conductivity vanishing rapidly at low frequencies, $\sigma_1(\omega)
\sim \omega^\alpha$ with $\alpha > 1$.  For $\Delta <2$, the Bose
metal will be unstable to superconductivity, but for small boson
hopping amplitude the transition temperature would be low.  In that
case, the optical conductivity above $T_c$ might still be well
described by the above power law form.

One may also directly calculate the non-linear {\sl dc} conductivity,
$\sigma(E_y,T) = \partial\langle I_y\rangle/\partial E_y$ as a function
of $E_y$ and $T$.  This is mathematically complicated, but formally
quite similar to the perturbative calculations of tunneling
conductance between parallel Luttinger liquids, as carried out e.g. in 
Ref.~\onlinecite{Carpentier}.  We forgo this calculation here for the
sake of brevity, quoting only the resulting scaling form
\begin{equation}
  \sigma(E_y,T) = A t^2 T^{2\Delta-3} {\cal G}(E_y/T),
\end{equation}
where ${\cal G}({\cal E}) \rightarrow 1$ as ${\cal E}\rightarrow 0$,
and ${\cal F}({\cal E})\sim {\cal E}^{2\Delta-3}$ for ${\cal E}\gg 1$.  This
implies in particular that the DC (linear response) conductivity at non-zero temperature
behaves as $\sigma(T) \sim A t^2 T^{2\Delta-3}$.  These results obtain
for the pure ring model, but will doubtless be altered somewhat in the
presence of impurity scattering.  This we leave for future study.

\section{\label{sec:instab}Instabilities of the Bose metal}

There are two classes of perturbations that one can add to the Bose
metal fixed point (described by the quadratic action ${\cal S}_0$)
that can potentially destabilize the phase.  The first are terms
involving the hopping of bosons.  When relevant, such boson hopping
terms ``stiffen'' the phase fluctuations of the bosons, leading to
off-diagonal long range order and a superconducting state.  As we
shall see, the perturbative relevance of the boson hopping terms is
determined by the Bose liquid parameters which enter into the fixed
point action which characterizes the Bose metal phase.  There are
regions of parameters where all such boson hopping terms are
irrelevant and the Bose metal phase is stable to such superconducting
perturbations, as we detail in Subsection A below.

The second class of perturbations involve hopping or motion of
vortices, conveniently expressed in the dual representation.  When
relevant, these perturbations signal a condensation of vortices which
typically leads to a breaking of translational symmetry and drives the
system into an incompressible insulating state.  The presence of these
instabilities and the precise form of the translational symmetry
breaking depend sensitively on the boson density, generally requiring
boson densities commensurate with the underlying lattice.  In
subsection B below we focus on half-filling ($\overline{n}=1/2$), and
study the nature of the resulting commensurate insulating states.

If the boson density is incommensurate with the lattice, on the other
hand, small vortex hopping terms are unimportant.  Provided the Bose
liquid parameters are in the regime where Boson hopping is likewise
irrelevant, the Bose metal exists as a completely stable critical
phase.  The gapless Bose metal is then a 2d analog of the stable 1d
Luttinger liquid.

\subsection{Boson hopping and superconductivity}

Consider then the stability of the Bose metal phase in the presence of
boson hopping operators.  To be general, we consider processes where
$q$ bosons hop along a displacement vector ${\bf s}$:
\begin{equation}
  {\cal L}_t = - \sum_{\bf r} \sum_{q,{\bf s}} t^{\bf s}_q
  \cos[q(\varphi_{\bf r} - \varphi_{{\bf r}+{\bf s}})] . 
\end{equation}
To assess the perturbative relevance of such processes, we compute the
two-point function of the tunneling operators,
\begin{equation}
{\cal T}_q^{\bf s}({\bf r}) = \cos[q(\varphi_{\bf r} - \varphi_{{\bf r}+{\bf s}})],
\end{equation}
using the Gaussian action for the Bose metal.  The row/column
symmetries in the Bose metal phase greatly constrain the spatial
dependence of these correlators.  We consider first the special class
of tunneling operators which hop bosons along the $x$ or $y$ axes with
${\bf s} = (m,0)$ or ${\bf s}=(0,m)$.  For this class of operators one
finds a power law decay both in time and in one spatial dimension:
\begin{equation}
  \langle {\cal T}_q^{m\hat{\bf y}}({\bf r},\tau) {\cal T}_q^{m\hat{\bf 
      y}} ({\bf 0},0) \rangle_0 = \delta_{y,0} [x^2 + v^2
  \tau^2]^{-\Delta_{qm}}  , 
\end{equation}
with scaling dimension,
\begin{equation}
  \Delta_{qm} = \frac{q^2}{2\pi} \int_{-\pi}^\pi \frac{dk_y}{2\pi}
  \sqrt{\frac{{\cal U}(0,k_y)}{{\cal K}(0,k_y)}}
  \frac{\sin^2(mk_y/2)}{|\sin(k_y/2)|}  =\frac{q^2}{2} \eta(m).
\end{equation}
Notice that these scaling dimensions vary as $q^2$, and will generally
increase (slowly -- $\Delta_{qm} \sim [q^2/(2\pi^2)] \sqrt{{\cal
    U}_0/{\cal K}_0}\ln |m|$) with hopping distance $m$.  As expected
these scaling dimensions increase with $U/K$.

For the remaining boson tunneling operators with vector ${\bf s}$
connecting two different rows and two different columns, the
correlator is further constrained by the row/column symmetry, being
spatially {\it local},
\begin{equation}
  \langle {\cal T}_q^{\bf s}({\bf r},\tau) {\cal T}_q^{\bf s}({\bf 0},0)
  \rangle_0 = G(\tau) \delta_{{\bf r},{\bf 0}} \qquad {\rm for}\,
  s_xs_y\neq 0, 
\end{equation}
with $G(\tau) \sim |\tau|^{-2 \Delta_q^{\bf s}}$.  

General renormalization group reasoning implies that operators are
irrelevant about a given fixed point when the associated scaling
dimension exceeds the space-time dimension, $\Delta > D = d+1$, with
dynamical exponent $z=1$.  But due to the constrained form of the
above correlators which only exhibit a power law decay in a reduced
set of space-time dimensions, $D_{red}$, one expects that the
condition for irrelevance should be modified to be $\Delta > D_{red}$.
Thus, when $\Delta_{q,m} > 2$ and $\Delta_q^{\bf s} > 1$, one expects
that all of the boson hopping operators should be unimportant as one
scales down in energy, and the Bose metal phase will be stable.  This
can always be achieved (in principle) by increasing the ratio of
$U/K$.

It is tempting to strengthen this expectation by constructing an
explicit renormalization group transformation, but this is somewhat
problematic due to the peculiar existence of zero energy states at
both $k_x=0$ and $k_y=0$ in the Bose metal phase.  One could try to
integrate out gapped modes away from the zero energy ``cross'' in the
Brillouin zone, and then successively integrate out ``shells'' of modes
pinching down onto the cross.  A difficulty arises, however, in the
rescaling transformation of the momenta, because the range of both
$k_x$ and $k_y$ -- the interval $[-\pi,\pi]$ -- is not invariant.
Similar but less severe difficulties are encountered when one tries to
implement a momentum shell RG procedure for a 2d Fermi surface, due to
the possible modifications of the shape/size of the Fermi surface.  It
might be possible to circumvent this difficulty along the lines of
Shankar {\sl et.  al.}\cite{Shankar}, or by an RG procedure in
frequency space.\cite{unpub}\  Some insight can be gleaned by ignoring
the zero modes along the $k_x=0$ axis, and considering a 1d RG
transformation where the integration is over a shell of momenta in
$k_x$ for all $k_y \in [-\pi,\pi]$ and frequency, and then rescaling
both $k_x$ and $\omega$ (ie. with $z=1$) but not $k_y$.  The resulting
perturbative (linearized) RG equation for the Boson hopping amplitude
in the $y-$direction, $t_{q,m} \equiv t^{m\hat{\bf y}}_{q}$ is then 
simply,
\begin{equation}
  \partial_\ell t_{q,m} = (1+z - \Delta_{q,m})  t_{q,m} ,
\end{equation}
with $z=1$ and $\Delta_{q,m}$ the scaling dimension given explicitly
above.  This argument indeed supports the expectation that weak Boson
hopping will be irrelevant provided $\Delta_{q,m} > D_{red}=2$.  In
the absence of a fully controlled 2d RG procedure, we can verify that
this conclusion in fact correct by resorting to perturbation theory.
In Appendix~\ref{sec:relevance}, we do just this by formally computing corrections to
the two-point function, $\langle e^{i\varphi_{\bf r}(\tau)}
e^{-i\varphi_{\bf r}(0)} \rangle$ perturbatively in powers of the
Boson hopping $t_{q=1,m=1}$.  Stability of the Bose metal phase
requires that the long-time behavior be unmodified from that
calculated within the Gaussian theory.  Carrying out this expansion to
second order in $t_{1,1}$, we show explicitly that this is in fact the
case provided $\Delta_{1,1} > 2$, thereby confirming the expected
result.

When the scaling dimension of a single boson hopping term is
sufficiently small, $\Delta_{1m} < 2$, one expects the Bose metal
phase to be unstable to a superfluid state in which the bosons
condense, with $\langle e^{i\varphi} \rangle \ne
0$.\cite{bosecondensenote}\ If the original lattice Bosons are
supposed to represent the Cooper pairs (say in a model of the
cuprates), then this will of course be the superconducting phase.

\subsection{Vortex hopping and insulating states}

\subsubsection{Stability}

We next consider the effects of the various non-linear interactions
involving the vortex operators in Eq.~(\ref{eq:onsite}) and
(\ref{eq:twosite}), which can potentially destabilize the Bose metal
phase.  At a generic density, for which $\bar{n}$ is irrational, all
these cosine terms are {\sl oscillatory}, and, if weak, cannot lead to 
any long-wavelength divergences.  Hence we expect that away from
commensurate densities, and provided $\Delta_{qm}>2$ for all $q,m$ so
that boson tunneling is irrelevant, the Bose metal is a stable zero
temperature phase of matter.  

For rational densities, some of the vortex operators will be
non-oscillatory, and must be considered more carefully.  In
particular, for very commensurate boson densities (i.e. $\bar{n}$
equal to a small-denominator rational fraction), it is not apriori
obvious whether the Bose metal can even in principle be stable to both
boson hopping and vortex operators.  To demonstrate the issues, we
present a stability analysis for the special set of rational fillings,
$\bar{n}=1/z$, where $z$ is a positive integer.  

At $\bar{n}=1/z$, the on-site terms take the form,
\begin{equation}
  \label{Lv}
  {\cal L}_v = - \sum_{\bf r} \sum_{q=1}^\infty
  \upsilon_{2q}\cos(2q\vartheta + \frac{2\pi q}{z}xy) .
\end{equation}
Note that although for $q \neq j z$, with integer $j\geq 1$, these
terms are spatially-varying, they are not wholly oscillatory, in the
sense that they all have a non-zero spatial average for constant 
$\vartheta$, e.g. for prime $z$, and $q\neq j z$, $[\cos (\frac{2\pi
  q}{z} xy)]_{xy} = 1/z$.  Hence even for $q\neq j z$, they cannot be
argued away 
simply on the grounds that they are oscillatory. 
Instead, to assess the importance of these perturbations, we again
consider the two-point function of these operators evaluated with the
Gaussian Bose metal action,
\begin{equation}
  G^{2q}_\vartheta ({\bf r},\tau) = \langle e^{i2q\vartheta_{\bf r}(\tau)}
  e^{-i2q\vartheta_{\bf 0}(0)} \rangle_0 = \delta_{{\bf r},{\bf 0}} F(\tau) ,
\end{equation}
with $F(\tau) \sim e^{-q^2\sqrt{{\cal K}_0/{\cal U}_0}\ln^2(\tau)}$ at
large times.  Due to the dual row/column symmetry this correlator is
spatially {\it local}, and is also ``short-ranged'' in time, vanishing
faster than any power law.  The associated scaling dimension is thus
infinite, and these operators are strongly irrelevant and will not
destabilize the Bose metal.  However, as we shall see, they will play
an important role in determining which of the various insulating
phases is selected when the Bose metal is driven unstable by the
two-site terms.

When the two-site terms in Eq.~(\ref{eq:twosite}) 
are weak, we can drop those that are spatially oscillating
and focus on the rest which take the form of
vortex hopping and creation terms:
\begin{equation}
{\cal L}_w = - \sum_{{\bf r},a,\pm} w_{a,\pm} {\cal O}_{a,\pm}({\bf r}) ,
\end{equation}
with operators, ${\cal O}_{a,\pm}({\bf r}) = \cos(2q\vartheta_{\bf r}
\pm 2q^\prime \vartheta_{\bf r + \bf s})$, where ``a'' labels the
various values of integers $q,q^\prime$ and the hopping vector ${\bf
  s}$.  Again, to establish the perturbative relevance of such terms
in the Bose metal phase, we evaluate the two-point correlators with
the Gaussian fixed point action. For the operators
${\cal O}_+$, the dual row/column symmetry is especially restrictive
and we find that the associated correlator is once again spatially
{\it local}:
\begin{equation}
\langle   {\cal O}_{a,+}({\bf r},\tau) {\cal O}_{a,+}({\bf 0},0) \rangle_0
= F(\tau) \delta_{{\bf r},{\bf 0}}   ,
\end{equation}
and ``short-ranged'' in time with $F(\tau) \sim
e^{-c_{a,q}ln^2(\tau)}$ (with constant $c_{a,q}$), so that like the
on-site terms, these operators cannot destabilize the Bose metal.
Similar behavior is found for ${\cal O}_-$, except for the special
class of operators with $q=q^\prime$, which correspond physically to
$2q$ vortices hopping along a vector ${\bf s}$.  Of these, except for the
special case with ${\bf s}=(m,0)$ and ${\bf s}=(0,n)$, the two-point
function is again spatially local, although it is now a power law in
time, $F(\tau) \sim |\tau|^{-2\Delta}$.  While such operators could
potentially destabilize the Bose metal if the power $\Delta <1$, they
will generally be less singular than the remaining class of operators
with vortices hopping along the $x$ or $y$ axes:
\begin{eqnarray}
{\cal L}_w = &-& w_{2q,m} \sum_{xy} \sum_{qm=jz} \lbrace
\cos[2q(\vartheta_{xy} - \vartheta_{x+m,y})] 
\nonumber\\
&+& \cos[2q(\vartheta_{xy} - \vartheta_{x,y+m})]\rbrace  .
\end{eqnarray}
The operators with $qm \neq jz$ (with integer $j$) are truly oscillatory
(i.e. have zero spatial average for constant $\vartheta$) at
$\bar{n}=1/z$ and have been dropped.  We expect the coefficients
$w_{2q,m}$ to be positive (with the minus sign out front), since these
operators will be generated at second order from the irrelevant
on-site terms: $w_{2q,m} \sim \upsilon_{2q}^2$.

For this last class of operators
the two-point function is delta-correlated in one-spatial direction
but a power law in the other and in time:
\begin{equation}
\langle {\cal O}^y_{2q,m}({\bf r},\tau) {\cal O}^y_{2q,m}({\bf 0},0) \rangle_0
\sim \delta_{y,0} (x^2 + v^2 \tau^2 )^{-\Delta^v_{2q,m}}  ,
\end{equation}
where 
${\cal O}^y_{2q,m}(x,y) = \cos[2q(\vartheta_{xy} - \vartheta_{x,y+m})]$.
The associated scaling dimension is given in terms of the Bose liquid parameters:
\begin{equation}
\Delta^v_{2q,m} = 2 q^2 \int_{0}^\pi dk_y\, \sqrt{\frac{{\cal K}(0,k_y)}{{\cal U}(0,k_y)}}  \frac{\sin^2(mk_y/2)}{\sin(k_y/2)} =2q^2{\eta_v}(m) .
\end{equation}
As for the boson hopping operators, the vortex hopping scaling
dimensions vary as $q^2$, and also increase slowly with increasing
hopping distance, $m$.  In general, we expect that for large $z$ the
most relevant of these will be $w_{2,z}$, with scaling dimension
$\Delta^v_{2,z}=2{\eta_v}(z) \sim 2\sqrt{{\cal K}_0/{\cal U}_0} \ln |z|$,
so for sufficiently large $z$ this can be rendered arbitrarily large,
and hence {\sl all} the vortex hopping terms can be made irrelevant
{\sl for any ${\cal U}({\bf k}), {\cal K}({\bf k})$}.  Thus for
sufficiently large $z$, such that $\Delta_{2q,m} > D_{red}=2$ for all
$qm = j z$, there is certainly a domain of stability for
the Bose metal phase (i.e. where the Bose liquid parameters are
further tuned to make all hopping terms irrelevant).  

For small $z$, however, this is not clear.  Indeed, it is
straightforward to show that for $z=2$ (i.e. half-filling,
$\bar{n}=1/2$), a choice of {\sl uniform} functions ${\cal U}({\bf k})
= {\cal U}_0, {\cal K}({\bf k})={\cal K}_0$ does not lead to a stable
regime.  In particular, in this case, the leading vortex hopping
operator has scaling dimension $\Delta^v_{2,2}= 2{\eta_v}(2)=4\sqrt{{\cal
    K}_0/{\cal U}_0}$, while the leading boson hopping operator has
scaling dimension $\Delta_{11}=\frac{1}{\pi^2}\sqrt{{\cal
    U}_0/{\cal K}_0}$.  Hence $\Delta^v_{22} \Delta_{11}=4/\pi^2$,
which implies ${\rm Min}\, (\Delta_{11}, \Delta^v_{22}) \leq 2/\pi <
2$, so that at least one or the other operator is relevant.  We have
not determined whether a stable Bose metal phase might be possible at
half filling when ${\cal U}({\bf k}), {\cal K}({\bf k})$ are
momentum-dependent.

\subsubsection{Instabilities}

As indicated above, for small $z$, either a vortex or boson hopping
instability may be inevitable.  In any case, it is interesting to
study the nature of the state resulting from relevant vortex hopping
terms.  Here, we briefly study the nature of the resulting phase for
half-filling, taking into account the presence of the two most
potentially relevant operators, with $q=1,m=2$ and $q=2,m=1$.  In
general, at half-filling, there are Bose liquid parameters for which
both $w_{22}$ {\sl and} $t_{1,1}$ are relevant, and more complex
behavior may well occur in this regime.  We will however neglect boson
hopping completely, as appropriate for large ${\cal U}/{\cal K}$.  

To this end, let us assume that $w_{2,2}$ is the {\sl most} relevant
operator, with $\Delta^v_{2,2}<2$.  We also assume that $w_{4,1}$ is
the next most relevant (it could be irrelevant, but still the {\sl
  least} irrelevant remaining operator).  Then provided all bare
couplings are small, we imagine integrating down in energy until
$w_{2,2}$ becomes comparable to the Gaussian terms in ${\cal S}_0$.
This requires $w_{2,2}\Lambda^{\Delta^v_{2,2}} \sim {\cal
  U}(\pi,0)\Lambda^2$, which always occurs for sufficiently small
$\Lambda$ with $\Delta^v_{2,2}<2$.  At this point it seems appropriate
to minimize the potential
$-w_{2,2}[\cos(2\vartheta_{xy}-2\vartheta_{x+2,y})+
\cos(2\vartheta_{xy}-2\vartheta_{x,y+2})]$, simultaneously with ${\cal
  S}_0$.  This minimization somewhat underestimates fluctuation
effects, which will be commented upon later.  The most general form
for $\vartheta_{xy}$ which minimizes the $w_{2,2}$ term and keeps the
Gaussian action small can be written
\begin{equation}
  \vartheta_{\bf r} = \Theta_0({\bf r}) + (-1)^x \Theta_1({\bf r}) +
  (-1)^y \Theta_2({\bf r}) + \frac{\pi a x}{2} + \frac{\pi b y}{2},
  \label{eq:w22decomp} 
\end{equation}
where $\Theta_{0/1/2}({\bf r})$ are slowly-varying functions of
space.  Inserting this expression for $\vartheta$ into the action
including ${\cal S}_0$ and the $w_{2,2}$ term only gives  
\begin{eqnarray}
  {\cal
  S}^{\rm eff}_0 & = & \int_{{\bf r}\tau} \bigg\{
  \frac{1}{\pi^2{\cal K}_0} (\partial_\tau\Theta_0)^2 +
  \frac{1}{\pi^2{\cal K}_\pi}
  [(\partial_\tau\Theta_1)^2+(\partial_\tau\Theta_2)^2] \nonumber \\
  & + & \frac{A_-}{2} (\nabla \Theta_0)^2 + \frac{A_-}{2}(\partial_x\Theta_1)^2+
  \frac{A_+}{2}(\partial_y\Theta_1)^2 \nonumber \\
  & + & \frac{A_+}{2}(\partial_x\Theta_2)^2+
  \frac{A_-}{2}(\partial_y\Theta_2)^2\bigg\}, \label{eq:heightmodel}
\end{eqnarray}
where ${\cal K}_\pi={\cal K}(\pi,0)$, $A_+ = 8 {\cal U}(\pi,0)/\pi^2$,
$A_- \sim w_{22} \Lambda^{\Delta^v_{2,2}} \sim
w_{2,2}^{2/(2-\Delta^v_{2,2})} [{\cal
  U}(\pi,0)]^{-\Delta^v_{2,2}/(2-\Delta^v_{2,2})}$, where $\Lambda$ is
the reduced low energy momentum cutoff.  Note that we are not
rescaling any fields or coordinates in this schematic RG treatment,
just taking into account the fluctuation corrections to $w_{2,2}$.
Thus the low-energy continuous fluctuations around the minima are
described by three massless fields $\Theta_{j}$ ($j=0,1,2$), which
unlike $\vartheta$ in the absence of $w_{2,2}$ have an ordinary
``relativistic'' dispersion.  In addition to these continuous
variables, the discrete degeneracy of distinct minima allowed by
periodicity of the cosine is indexed by the integers $a,b$.

Having taken into account $w_{2,2}$ already in
Eq.~(\ref{eq:heightmodel}), we next include the effects of a weak,
renormalized $w_{4,1}$ coupling, which after renormalization is of
order $w_{4,1}^R \sim w_{4,1} \Lambda^{\Delta^v_{4,1}} \sim
w_{4,1}(w_{2,2}/{\cal U}(\pi,0))^{\Delta^v_{4,1}/(2-\Delta_{2,2}^v)}$.
If all the bare non-linear couplings are small, then after
renormalization, this will be the {\sl largest} remaining term, which
is why we treat it next.  Inserting the decomposition of
Eq.~(\ref{eq:w22decomp}) into the four vortex hopping term, one
obtains
\begin{equation}
  {\cal S}_1^{\rm eff} = -w_{4,1}^R \int_{{\bf r}\tau} \left[\cos
    8\Theta_1 + \cos 8\Theta_2\right], \label{eq:seff1}
\end{equation}
assuming small gradients of $\Theta_{1/2}$, as mandated by
Eq.~(\ref{eq:heightmodel}).
Eqs.~(\ref{eq:heightmodel},\ref{eq:seff1}) describe a 3d ``height''
model for $\Theta_1,\Theta_2$.  As is well known in such models, the
fluctuations of the free scalar fields $\Theta_1,\Theta_2$ are {\sl
  bounded}, so that the cosine terms in Eq.~\ref{eq:seff1}\ are always
relevant and pin these fields to integer multiples of $\pi/4$, which we
can take, without loss of generality, to be zero.  Thus the net effect of
$w_{4,1}$ is to leave only $\Theta_0$ as a low-energy mode.

Setting therefore $\Theta_1=\Theta_2=0$ in the pinned (``flat'')
phase, we have
\begin{equation}
e^{2i\vartheta_{x,y}} = e^{2i \vartheta_0 } (-1)^{ax+by}    .
\end{equation}
Here, since the fluctuations of $\Theta_0$ are
likewise bounded, we have replaced it by its average (zero mode) value
$\Theta_0 \rightarrow \vartheta_{\bf 0}$.  This average can be
determined by minimizing the on-site Lagrangian in Eq.~(\ref{Lv}).
In principle, the parameters in Eq.~(\ref{Lv}) should also be
renormalized, e.g. $\upsilon_{2q} \rightarrow \tilde{\upsilon}_{2q} \sim
\upsilon_{2q} \exp[-q^2 c_q \ln^2 (w_{2,2}/{\cal U}(\pi,0))]$.
This gives,
\begin{equation}
  e^{2i \vartheta_{\bf 0} } = (-1)^{ab}    ,
\end{equation}
for $\tilde{\upsilon}_4 > -\tilde{\upsilon}_2/8$ (with
$\tilde{\upsilon}_2 >0$ assumed) and 
\begin{equation}
  e^{2i \vartheta_{\bf 0} } = (-1)^{ab} [
  (\tilde{\upsilon}_2/8|\tilde{\upsilon}_4|) \pm i \sqrt{1 -
    (\tilde{\upsilon}_2/8\tilde{\upsilon}_4)^2} ]    , 
\end{equation}
for $\tilde{\upsilon}_4 < -\tilde{\upsilon}_2/8$.

The spatial orderings that are implied by these mean field solutions
follow readily from the expressions in Eq.~(\ref{eq:opcontentn},\ref{eq:opcontente}) 
relating the bare boson density and plaquette energy density
to the ``low-energy'' field $\vartheta$.  In the former case above,
the boson density is uniform since $\sin(2q \vartheta_{x,y})=0$,
whereas there is a plaquette energy density wave with,
\begin{equation}
\epsilon_{x,y} \sim c^\epsilon_2 (-1)^{(x+a)(y+b)}  .
\end{equation}
The four states with $a,b=0,1$ correspond to the four plaquette
density wave states of the original lattice Boson model in which
one out of every four plaquettes is resonating more strongly.

When $\tilde{\upsilon}_4 < -\tilde{\upsilon}_2/8$, on the other hand,
since $\sin(2\vartheta) \ne 0$ 
the plaquette energy density wave states (with reduced order
$c^\epsilon_2 \rightarrow 
 (\tilde{\upsilon}_2/8|\tilde{\upsilon}_4|) c^\epsilon_2$) are
 co-existing with a charge-density-wave 
(CDW) state ordered at $(\pi,\pi)$:
\begin{equation}
\delta n_{x+ \frac{1}{2} , y + \frac{1}{2} } \sim  \pm c^\rho_2 
\sqrt{1 - (\tilde{\upsilon}_2/8\tilde{\upsilon}_4)^2} (-1)^{x+y}   .
\end{equation}
In the limit in which $\tilde{\upsilon}_4 \rightarrow - \infty$,
the amplitude of the plaquette density wave vanishes, leaving only
CDW order.  We believe that an absence of a pure CDW state
more generally is an artifact of the mean-field treatment
being employed, which ignores all fluctuations in the $\vartheta$ field.
In particular, one can imagine domain walls forming
between domains of different plaquette-density-wave order,
which cost an energy $E_{wall} \sim w_{2,2}
(\tilde{\upsilon}_2/\tilde{\upsilon}_4)^2$ 
for small $w_{2,2}$ and large $|\tilde{\upsilon}_4|$.  When this
energy is small, 
fluctuations will undoubtedly disorder the plaquette-density-wave
order, leaving the pure CDW.

\begin{figure}
\begin{center}
\vskip-2mm
\hspace*{0mm}
\centerline{\fig{2.8in}{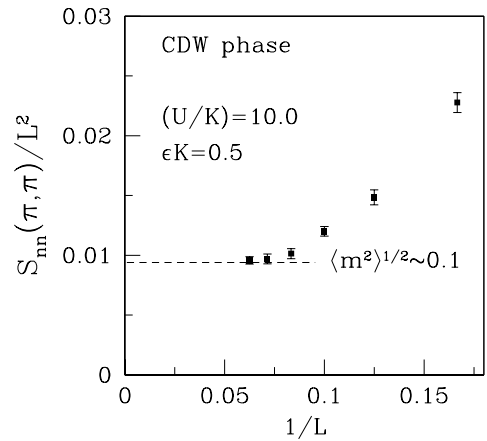}}
\vskip-2mm
\caption{Finite size scaling of the density structure factor peak
  at $(\pi,\pi)$, well within the incompressible phase. The staggered
  magnetization $m=\langle m^2\rangle^{1/2}$ clearly saturates to a
  finite value $1/L \to 0$, indicating a non-zero $(\pi,\pi)$ charge
  density wave order parameter.}
\label{fig:cdwmoment}
\end{center}
\end{figure}

\begin{figure}
\begin{center}
\vskip-2mm
\hspace*{0mm}
\centerline{\fig{2.8in}{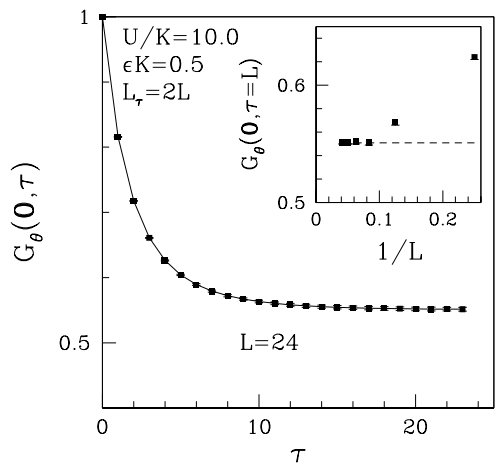}}
\vskip-2mm
\caption{Vortex correlation function $G_\theta({\bf 0},\tau)$
in the incompressible CDW phase. The correlation
function saturates as seen from the finite size scaling plot in
the inset indicating that the vortices condense,
leading to a conventional insulating state.}
\label{fig:cdwtau}
\end{center}
\end{figure}

Our quantum Monte Carlo simulations on the boson ring model in the
dual representation reveal that in the large $U$ limit with $U/K >
(U/K)_c \approx 2.5$, the Bose metal phase is unstable to the
formation of a CDW state.  This is apparent in
Fig.~\ref{fig:cdwmoment} which shows that the density structure factor
peak at $(\pi,\pi)$ grows as $L^2$, indicating the presence of
long-ranged CDW order.  This order is, however, weak, with a staggered
magnetization $m \approx 0.1$ (see Fig.~\ref{fig:cdwmoment})), much
smaller than the classical value $m=0.5$.  It is also instructive to
examine the vortex two-point function in the CDW state, which is shown
in Fig.~\ref{fig:cdwtau}.  Notice that the vortex correlation function
is long-ranged indicating a ``vortex condensation'' ($\langle
e^{i\theta} \rangle \ne 0$), as expected in such a conventional
insulating state.

Both the plaquette and charge density wave states will be insulators,
with a charge gap.  This follows since in both cases the field
$\vartheta$ is ``pinned'' by the cosine potentials and is not
fluctuating.  Since the bare boson density is $ n = \Delta_{xy}
\theta/ \pi$, adding a particle at the origin can be achieved by
shifting $\theta_{xy} \rightarrow \theta_{xy} + \pi$ for all $x,y >0$.
This will cost a finite amount of energy (coming from the plaquette at
the origin) when the field $\vartheta$ is pinned.

\section{Exact Wavefunction}

In this Section we consider a variant of the boson ring Hamiltonian
which allows us to obtain an exact zero energy wavefunction
when the couplings are carefully tuned.
We then perturb away from the soluble point, using the 
exact wavefunction to compute properties of the adjacent quantum phase.
Specifically, we find a translationally invariant fluid phase
with a finite compressibility, behavior consistent with both a superfluid
and the Bose metal.  But a calculation of the boson tunneling gap
in a finite system shows a $1/L$ scaling as expected for the Bose metal, and inconsistent with the $1/L^2$ dependence of a 2d superfluid.
Thus, we confidently conclude that the Bose metal phase
exists over a finite portion of the phase diagram adjacent to the soluble point.

The model we consider is a ``hard-core'' version of the boson ring
model, in which only zero or one boson is allowed per site, $b^\dagger
b = 0,1$.  In the
usual way we represent the hard-core bosons as Pauli matrices,
\begin{eqnarray}
  b^\dagger & = & \sigma^+, \\
  b & = & \sigma^-, \\
  n=b^\dagger b & = & \frac{1}{2}(1+\sigma^z),
\end{eqnarray}
where $\vec\sigma$ is the standard vector of
Pauli matrices, and $\sigma^\pm = \frac{1}{2}(\sigma^x \pm i
\sigma^y)$.  The Hamiltonian we consider consists of two terms:
\begin{eqnarray}
  \label{eq:spin12}
 H_{1/2} & = &   \sum_{xy}\bigg\{ -J_4 ( \sigma^+_{x,y} \sigma^-_{x+1,y}
  \sigma^+_{x+1,y+1} \sigma^-_{x,y+1} +  {\rm h.c.} ) \nonumber \\ && + u_4
  \hat{P}_{\rm flip}(x,y)\bigg\}   , \label{eq:Sring}
\end{eqnarray}
where
\begin{equation}
  \hat{P}_{\rm flip}(x,y) = \frac{1}{16}\sum_{\alpha=\pm 1}
  \prod_{\beta,\gamma=\pm
    1} \left (1 +\alpha\beta\gamma
    \sigma^z_{x+\frac{1}{2}+\frac{\beta}{2},y
      +\frac{1}{2}+\frac{\gamma}{2}}\right).
  \label{eq:Pflip}
\end{equation}
The first term with $J_4>0$ in
Eq.~(\ref{eq:Sring}) is the hard-core analog of the ring hopping term
proportional to $K$ in
the rotor model, Eq.~(\ref{eq:rotorham}).
The operator $\hat{P}_{\rm flip}(x,y)$ is a projection operator onto the
two flippable configurations of the square plaquette whose lower-left
corner is at the site $(x,y)$.   For $u_4>0$, this term competes with the ring term by
{\sl disfavoring}  configurations with flippable plaquettes.

Remarkably, the ground state of $H_{1/2}$ can be found exactly for the
special Rokhsar-Kivelson (RK) point $u_4=J_4$, following a general
construction in the spirit of the Rokhsar-Kivelson point of the square
lattice quantum dimer model,\cite{RK}\ and employed more recently in
Ref.~\onlinecite{BGF}\ for a similar spin model on the Kagome lattice.
The solution can be seen by rewriting $H_{1/2}$ as
\begin{equation}
  H_{1/2} =  \sum_{xy}
  \hat{P}_{\rm flip}(x,y)\lbrace J_4\big(1-\!\!\!
  \prod_{u,v=0,1}\sigma^x_{x+u,y+v}
    \big)  - v \rbrace,
    \label{HamRK}
\end{equation}
where $v=J_4-u_4$ measures the deviation from RK point.  For $v=0$, an
obvious (zero energy) ground state of $H_{1/2}$ is the fully-polarized
state
\begin{equation}
  |0\rangle = \prod_{x,y}|\sigma^x_{xy}=1\rangle.
  \label{eq:uniformstate}
\end{equation}
A useful alternate representation follows by rewriting
$|\sigma^x=1\rangle= \frac{1}{\sqrt{2}} (|\sigma^z=1\rangle +
|\sigma^z=-1\rangle)$, and expanding out the direct product,
\begin{equation}
  |0\rangle = \frac{1}{2^{N/2}}\sum_{\{\sigma_{xy}=\pm 1\}} \prod_{xy}
  |\sigma^z_{xy}=\sigma_{xy}\rangle,
\end{equation}
which demonstrates that $|0\rangle$ is a uniform real
superposition of all configurations in the $S^z$ (boson number) basis.

This state, however, has
uncertain $\sigma^z$ (boson number), and so can be decomposed into many
distinct ground states by projection.  In particular, at a given
average $\langle \sigma^z_{xy} \rangle = 2\langle  n_{xy}\rangle-1=  m$, we
may project onto
spatially uniform
states according to
\begin{equation}
  |0;m\rangle = \frac{1}{\sqrt{\tilde{Z}}}\prod_{x,y=1}^L \hat{P}_{x}(m)
  \hat{P}_y(m) |0\rangle,
\end{equation}
where $\tilde{Z}$ is a normalization constant,
since the row/column projection operators,
\begin{eqnarray}
  \hat{P}_x(m) & = & \int_0^{2\pi} \frac{d\theta}{2\pi} \exp\left[i\theta_x\sum_y
    \left(\sigma_{xy}^z-m\right) \right],  \nonumber \\
  \hat{P}_y(m) & = & \int_0^{2\pi} \frac{d\phi}{2\pi} \exp\left[-i\phi_y\sum_x
    \left(\sigma_{xy}^z-m\right) \right], \label{eq:projectors}
\end{eqnarray}
commute with $H_{1/2}$ and amongst themselves, reflecting the
conservation of boson number on each row and column by the ring
dynamics.

Of course, other {\sl non-uniform} ground states may be obtained by choosing
$m$ differently on different rows and columns.  This vast degeneracy
signals a pathology of the RK point, which is in fact at the boundary
of a first-order transition to a phase-separated ``frozen'' regime,
for $v<0$.  We therefore focus instead on the behavior for
infinitesimal $v>0$, which splits this degeneracy.  For boson
densities near half-filling, i.e. $|m| \ll 1/2$, we expect that the
uniform states will be favored energetically as $v$ is increased to
slightly positive values, since these states have in this density
regime more flippable plaquettes (see below).  For larger $|m|$, this
assumption is certainly violated, however, since e.g. at very low
boson density the ring moves clearly do not connect all possible
configurations.  At $m=0$, however, the set of uniform configurations
does form a single ergodic component under the ring move (as can be
straightforwardly shown numerically\cite{unpublished}\ and probably
argued analytically).  So we expect the set of uniform states to be an
adequate description for small $|m|$.

To make contact with the Bose metal fixed point description, we
calculate some energetic properties for infinitesimal positive $v$
using first-order perturbation theory.  From the splitting of the
different projected states at $O(v)$, we calculate two quantities: (1)
the ground state energy density $\epsilon(m)=E(m)/L^2$ as a function of boson
density $n=(m+1)/2$, from which we obtain the compressibility
$\kappa(m)$, and (2) the ``single-particle'' gap $\Delta_1(m,L)$ for a
finite-size $L\times L$ system, essentially the addition energy for a
boson {\sl onto a particular row and column} in the grand canonical
ensemble, defined more precisely below.

Consider first the ground state energy density.  For the projected
state at magnetization $m$, the first order energy shift is
\begin{eqnarray}
  E(m) & = & -v \sum_{xy}\langle 0;m|\hat{P}_{\rm flip}(x,y)|0;m
  \rangle \nonumber \\
  &  &  \hspace{-0.5in}= -\frac{v}{\tilde{Z}} \sum_{xy}\langle 0| \hat{P}_{\rm flip}(x,y)
  \prod_{x,y=1}^L \hat{P}_{x}(m) \hat{P}_y(m) |0\rangle,
\end{eqnarray}
where the latter equality follows from the fact that $\hat{P}_{\rm
  flip}$ is diagonal in the $\sigma^z$ basis and hence commutes with
the projection operators.  Using Eq.~(\ref{eq:uniformstate}), one readily
sees that the energy shift can be rewritten in a form reminiscent of
{\sl classical} statistical mechanics,
\begin{equation}
  E(m) = -v \sum_{xy} \left\langle P_{\rm
    flip}(x,y;\{\sigma\})\right\rangle_{\{\sigma_{xy}\}}
\end{equation}
where
\begin{equation}
  \langle {\cal O}\rangle_{\{\sigma_{xy}\}} =
  \frac{1}{Z}\sum_{\{\sigma_{xy}=\pm 1\}} {\cal O}\prod_{x,y=1}^L
  P_{x}(m;\{\sigma\}) P_y(m;\{\sigma\}). \label{eq:Ising}
\end{equation}
Here $P_{\rm flip}$, $P_x$ and $P_y$ are the classical functions obtained from
$\hat{P}_{\rm flip}$, $\hat{P}_x$, $\hat{P}_y$, respectively, by replacing the
operator $\sigma^z_{xy}$ by $\sigma_{xy}$, and $Z$ is chosen such that
$\langle 1 \rangle_{\{\sigma_{xy}\}} =1$.

We see that Eq.~(\ref{eq:Ising}) simply defines an expectation value
for an ``infinite temperature'' Ising model with a constrained
magnetization on each row and column.  By proceeding along similar
lines, one can calculate the energy shift for a state with an
extra boson on a single particular row and column, and hence obtain
$\Delta_1$.  Amusingly, both problems can be solved exactly using a
saddle-point technique (see Appendix~\ref{sec:saddle}).  The results
demonstrate that the constraints are ``nearly irrelevant'', and the
lattice gas behaves nearly as its unconstrained counterpart in the
large system limit.  In particular, the
energy density, as $L\rightarrow\infty$, becomes
\begin{equation}
  \epsilon(m) = -\frac{v}{8}(1-m^2)^2. \label{eq:energydensity}
\end{equation}
This can be easily understood by assuming that each site is completely
independent, and that the only effect of the
constraint is to determine the relative probabilities of the two spin
states, to wit $p_\uparrow=(1+m)/2, p_\downarrow =(1-m)/2$.  On a
given four-site plaquette, only the two (of 16 total) configurations in which the
spin alternates around the plaquette are flippable.  Hence the average
flippability per plaquette is $2 p_\uparrow^2 p_\downarrow^2 =
(1-m^2)^2/8$, in agreement with the result above.

The ``single-particle'' gap is somewhat less intuitive.  To be precise and
avoid ambiguities due to an unspecified chemical potential, we define
\begin{equation}
  \Delta_1(m) = \frac{1}{2}\left[E_1(m;+1)-2 E_1(m;0)
    +E_1(m,-1)\right],
  \label{eq:delta1def}
\end{equation}
where $E_1(m;\lambda)$ is the ground-state energy of a state in which
the number of bosons on row and column $x=y=1$ is increased by
$\lambda$ relative to the the uniform state at magnetization $m$:
\begin{eqnarray}
|0;m,\lambda\rangle & = & \frac{1}{\sqrt{Z'}}\hat{P}_{x=1}(m+2\lambda/L)
  \hat{P}_{y=1}(m+2\lambda/L)  \nonumber \\
  && \times\prod_{x,y=2}^L \hat{P}_{x}(m)
  \hat{P}_y(m) |0\rangle, \label{eq:extraproj}
\end{eqnarray}
and $E_1(m;\lambda)=-v\sum_{xy}\langle 0;m,\lambda|\hat{P}_{\rm flip}(x,y)|0;m,\lambda
  \rangle$.  One finds
\begin{equation}
  \Delta_1(m) = \frac{2 v}{L} (1-m^2) + O(1/L^2). \label{eq:opg}
\end{equation}

These two quantities can be compared to the general predictions
expected from our theory of the Bose metal phase.  Consider first the compressibility.  From
Eq.~(\ref{eq:energydensity}), one has $\mu = -d\epsilon/dn =
-2 d\epsilon/dm=v m(1-m^2)$, and hence
\begin{equation}
  \kappa = \frac{dn}{d\mu}=
  \frac{1}{2v}\frac{1}{1-3 m^2}. \label{eq:Isingcompress}
\end{equation}
Note that the compressibility diverges and becomes negative for
$1/\sqrt{3}<|m|<1$, indicative of an instability of the uniform state
well away from half-filling.  For $|m|<1/\sqrt{3}$, however, $\kappa$
is finite, consistent with the Bose metal phase.  Indeed,
from our general Harmonic description of the Bose metal phase
as in Eq. ??, we have $\kappa^{-1} = {\cal U}({\bf k} = 0)$.

In order to rule out a superfluid phase, we now compute the
single-particle gap, $\Delta_1(L)$, using the Bose metal Gaussian
fixed point theory to show that it varies as $1/L$ as in Eq.~(\ref{eq:opg}).
This gap can be extracted from the spatially local correlator,
\begin{equation}
G_{\varphi}(\tau) = \langle e^{i\varphi_{\bf r}(\tau)} e^{-i\varphi_{\bf r}(0)} \rangle ,
\end{equation}
evaluated in a  
{\sl finite $L\times L$ system}.
$ $From the spectral representation, one has, at zero temperature,
\begin{equation}
  \Delta_1(L) = \lim_{\tau \rightarrow \infty} - \frac{\ln
    G_\varphi(\tau)}{\tau}.
\end{equation}
Performing the Gaussian integral using the effective action in
Eq.~(\ref{Gaussianphi}), one has
\begin{equation}
  -\ln G_\varphi(\tau) = \int_{-\infty}^\infty \!
  \frac{d\omega}{2\pi} \frac{1}{L^2} \sum_{\bf k} \frac{{\cal U}({\bf
      k})}{\omega^2 + E_{\bf k}^2} \left( 1-e^{-i\omega\tau}\right),
  \label{eq:finitesizeint}
\end{equation}
where the wavevector sum is over inequivalent values with $k_x,k_y \in
(2\pi/L){\cal Z}$ in the first Brillouin zone, and
with the mode energy $E_{\bf k}$ given explicitly in Eq.~(\ref{modeenergy}).
As $\tau\rightarrow \infty$, the $\omega$ integral in
Eq.~(\ref{eq:finitesizeint}) is dominated by those terms for which
$E_{\bf k}=0$.  This occurs along the Bose surface, i.e. for $k_x=0$
or $k_y=0$.  In the finite size system there are $2L-1 \approx 2L$
such points (for $L\gg 1$), hence
\begin{eqnarray}
  -\ln G_\phi(\bbox{0},\tau) \sim \frac{2 \overline{\cal U} }{L}  \int_{-\infty}^\infty \!
  \frac{d\omega}{2\pi} \frac{1}{\omega^2} \left( 1-e^{-i\omega\tau}\right),
\end{eqnarray}
where
\begin{equation}
  \overline{\cal U}=\frac{1}{2}\left[ \int_{-\pi}^\pi \!
    \frac{dk_x}{2\pi} {\cal U}(k_x,0) + \int_{-\pi}^\pi \!
    \frac{dk_y}{2\pi} {\cal U}(0,k_y)\right],
\end{equation}
is the Bose-surface average of ${\cal U}({\bf k})$.  Thus, we finally have,
\begin{equation}
  \label{eq:Gdecay}
  -\ln G_\phi(\bbox{0},\tau) \sim \frac{ \overline{\cal U}|\tau|}{L},
\end{equation}
giving the $1/L$ behavior for the single-particle gap:
\begin{equation}
  \label{eq:spgap12}
  \Delta_1=\frac{ \overline{\cal U} }{L}.
\end{equation}
In a 2d superfluid, the single-particle gap is much smaller,
vanishing as $1/L^2$ in the large $L$ limit.

The agreement of the finite compressibility and $1/L$ scaling of
$\Delta_1$ between the soluble model and our fixed point theory
of the Bose metal strongly argues that
the ground state of $H_{1/2}$ is the Bose liquid for $0<v\ll 1$ and
$|m|<1/\sqrt{3}$.  If this postulate is correct, certain combinations of the Bose liquid function ${\cal U}({\bf k})$ can be obtained explicitly for the soluble model. 
In particular, we find
\begin{eqnarray}
  \label{eq:relations}
  {\cal U}({\bf k} =0) & = & 2 v(1-3m^2), \\
  \overline{\cal U}  & = & 2v(1-m^2)  .
\end{eqnarray}

\section{Discussion\label{sec:discuss}}

\subsection{Fractionalization, the $Z_2$ gauge theory and the
  high-$T_c$ cuprates}

In this paper we have described a novel phase of quantum matter, the
Bose metal, which we argue occurs in a class of square lattice boson
ring models with XY symmetry.  Having done so, it is reasonable to
reflect upon the context in which such models are physically
appropriate and the consequences perhaps observable.  As discussed
briefly in the introduction, the primary motivation for these models
comes from the high temperature cuprate superconductors.  Both the
remarkably high critical temperatures and the strange behaviors in the
``normal'' state of these materials motivated a number of theorists
early on to the radical suggestion that spin-charge separation might
underly these peculiarities.\cite{Anderson,KRS,LN,SenMPAF}\ In this picture,
the electron charge, liberated both from its spin and its Fermi
statistics as a ``chargon'' (or ``antiholon''), is relatively free to
Bose condense to form a superconducting state.  At higher
temperatures, or when the material is under-doped, the chargons and
spin-carrying fermionic ``spinons'' form a novel fluid, which would no 
doubt behave very differently from a normal metal.  

Subsequent theoretical and consequent experimental work has both
advanced the status of these ideas and posed a severe challenge to
their applicability to the cuprates.  A recent
formulation\cite{Z2A,Z2B}\ of interacting electrons in terms of
spinons and chargons minimally coupled to a $Z_2$ gauge field has
helped to support the concept of electron fractionalization with a
very concrete formal framework.  The essential elements of the $Z_2$
gauge theory are reviewed in Appendix~\ref{app:z2}.  As with earlier
$U(1)$ and $SU(2)$ gauge theory formulations\cite{U1A,U1B,U1C}, the
``spinons'' are taken as Fermions carrying the spin of the electron but
are electrically neutral.  The bosonic ``chargons'' carry the electrons
charge.  The $Z_2$ gauge theory provides a convenient phenomenology
for describing a fractionalized phase in which the spinons and
chargons are deconfined, and live as well-defined particle
excitations.  Because of its concrete formulation, and the virtue that
the $Z_2$ gauge theory has a well-understood confinement-deconfinement
phase transition, it is possible to calculate from it in a simple way
the qualitative (universal) properties of the fractionalized state.
One of the most fundamental of these properties is the existence of
gapped topological excitations\cite{vison1,vison2,vison3} called
``visons''.  The visons act as sources of $Z_2$ flux, and may be
thought of as remnants of unpaired superconducting vortices in the
fractionalized insulator\cite{Z2A}, which in turn is viewed as a
paired-vortex condensate\cite{BFN2}.  Understanding these excitations
led to a rather direct proposal for a ``vison trapping''
experiment,\cite{Z2B}\ designed to trap and detect visons by cycling
through the superconducting--normal transition.  Current
experiments\cite{visexp1,visexp2} apparently imply that the gap to
such vison excitations in the insulator, if it is non-zero at all, is
less than 190$K$.  Since the presence of a vison gap is a necessary
condition for the existence of a true spin-charge separated ground
state, this unnaturally low energy scale presents a difficult obstacle
to theories of fractionalization as applied to at least these
particular cuprate materials (samples of underdoped BSSCO and YBCO).

The spin-charge separation scenario is nonetheless extremely appealing
theoretically, and it is interesting to consider the possibility of
retaining some degree of this physics locally.  One may imagine for
simplicity an undoped model which interpolates between at one extreme
a conventional ``Hubbard-like'' insulator with gapped charge degrees
of freedom and Heisenberg-interacting spins, and at the other extreme
a fully fractionalized insulator.  The $Z_2$ gauge theory is
well-suited for this purpose.  The interpolation between the above two
limits is accomplished in this model by varying a coefficient $K$,
which controls the strength of fluctuations in the gauge field.  The
deconfined phase is obtained in the large $K$ limit, while we consider
here the first deviations away from $K=0$, which is deep within the
confined, Hubbard-like phase.  As we show in Appendix~\ref{app:z2}, 
when the gauge theory is deep within
it's confined phase (with $K$ small), the gauge fields
can be formally integrated out and one recovers a Hamiltonian
expressible in terms of electron operators and composites built from
the electron such as the spin operator, together with a Cooper pair
field.  In the spin sector, we find that the leading terms obtained in
this limit are an antiferromagnetic nearest-neighbor Heisenberg
exchange (for $K=0$) and {\sl a plaquette ring
  term},
\begin{equation}
H^s_\Box = J^s_{\Box} \sum_{\Box} {\cal R}^s_\Box  , \label{eq:plaqspin}
\end{equation}
with $J^s_\Box \sim O(K)$ ($J^s_\Box = K (t_s^4 + \Delta^4)/(h+U)^4$,
in terms of the parameters in Appendix~\ref{app:z2}).  Here, ${\cal
  R}^s_\Box$ is a plaquette ring operator defined in terms of the four
spins on a given plaquette\cite{oldring1,oldring2}:
\begin{eqnarray}
{\cal R}^s_\Box &=& \sum_{i<j=1}^4 {\bf S}_i \cdot {\bf S}_j + 4({\bf
  S}_1 \cdot {\bf S}_2 )  ({\bf S}_3 \cdot {\bf S}_4 ) \nonumber\\ 
&+&
 4({\bf S}_1 \cdot {\bf S}_4 )  ({\bf S}_2 \cdot {\bf S}_3 ) -  4
 ({\bf S}_1 \cdot {\bf S}_3 )  ({\bf S}_2 \cdot {\bf S}_4 ) . 
\end{eqnarray}
In the charge sector, to $O(K)$ one
finds the leading term
\begin{equation}
H^c_\Box = - J^c_\Box \sum_\Box {\cal R}^c_\Box   , \label{eq:plaqcharge}
\end{equation}
with $J^c_\Box \sim O(K)$ ($J^c_\Box = K t_c^4/(h+U)^4$, derived in Appendix~\ref{app:z2}).
Here, ${\cal R}^c_\Box$ is a ring operator in the charge sector
expressed simply in terms of Cooper pair operators,
$e^{i\phi_{\bf r}} \equiv b_{\bf r}^2$ as,
\begin{equation}
{\cal R}^c_\Box = \sum_{i<j=1}^4 \cos(\phi_i - \phi_j) + \cos(\Delta_{xy} \phi_{\bf r})   .
\end{equation}
Notice that the second term above is precisely of the form taken in
our starting Hamiltonian.  As $K$ is increased from zero, such
ring-exchange processes play more and more important roles.  We
thus view the ring Hamiltonian studied in this paper as a suitable
model that offers an intermediate ground between the fully spin-charge
separated scenario and conventional phases of matter.  

Should this bosonic ring model for the charge sector have some
relevance for the cuprates, what might be the consequences and
interpretation?  As we have seen in Sec.~\ref{sec:instab}, the ring
model sustains the Bose metal phase for sufficiently repulsive
interactions (large $U$) and {\sl incommensurate densities}.  Indeed,
as we saw in Sec.~\ref{sec:instab}, a simplistic estimate indicates
that there is no stable regime for the Bose metal at half-filling.
Hence, if we assume the strong interaction condition obtains, one
would expect an evolution from a conventional insulator with a charge
gap at half-filling to the Bose metal upon doping sufficiently away.
If one further presumes (as seems natural) that the generalized
``stiffness'' ${\cal K}$ increases with doping, then one would expect
further doping to lead to a superconducting (boson hopping)
instability.  This scenario thus naturally associates the Bose metal
with the pseudo-gap regime of the high-$T_c$ cuprates.  

\subsection{Extensions}

To determine if there is any truth to the above scenario requires
considerable extensions of the present work.  Most significantly, spin
and quasiparticle degrees of freedom are important components of the
high-$T_c$ materials, and should be incorporated into the description.
It will be interesting to consider interactions of the Fermionic
quasiparticles with the strongly fluctuating collective modes of the
Bose metal.  It seems to us quite likely that the Bose metal can
remain stable in the presence of these interactions, while perhaps at
the same time producing rather strong modifications of the fermionic
degrees of freedom.  In any case, it should be possible to consider
photoemission spectra and local electron tunneling density of states
in a model with the quasiparticles coupled to the Bose metallic modes.

We have also left a number of issues within the purely bosonic
description unanswered.  To understand transport measurements
generally will require an understanding of the effects of disorder.  
Several experiments attempting to access the ``normal'' state of the
cuprates employ large local or uniform magnetic fields to suppress
superconductivity\cite{Taillefer}.  Such magnetic fields 
indeed tend to suppress the
superconducting instabilities of the Bose metal, and one should
explore the ramifications of a Bose metallic description for e.g.
transport measurements in high fields.  

There are a number of other possible applications of the present
formalism to frustrated magnets with appreciable ring
exchange\cite{ringnum}.  Magnetic ring exchange processes are believed
important for instance in the Wigner crystal phase of the
two-dimensional electron gas\cite{UCLA,Ceperley} at $r_s \gtrsim 40$.
The effect of such processes are generally difficult to analyze, but a
perhaps illuminating approach may be to consider the easy-plane limit
of such spin ring models.  A straightforward calculation shows that
such a limit recovers XY ring models of the sort studied here
(representing the spins by hard-core bosons).  The plaquette duality
constructed here for the square lattice can be straightforwardly
generalized to other lattices, such as the triangular and Kagome
cases.  For the bosonic Kagome case, one can directly in this way
construct a dual formulation from which it is straightforward to
construct the visons and spinons of Ref.~\onlinecite{BGF}.  Whether
the spinons are confined or deconfined in the pure XY ring model on
this lattice can be easily established numerically using the dual
$\theta$ variables, and this investigation is underway.\cite{hermele}\ 
On the triangular lattice with four-site (parallelogram) ring
exchange, the plaquette duality shows that there is no stable metallic
state in this case, but a numerical approach will be necessary to
determine the ultimate nature of the ground state.

On the purely theoretical side, we have introduced the mathematics
necessary for a $2+1$ dimensional ``bosonization'' scheme, applied
here to ``bosonize bosons'' in terms of collective $\varphi,\vartheta$ 
modes.  We expect it should be possible to bosonize {\sl fermionic}
ring models in a similar way.  Consider {\sl spinless} fermions $\{
f_i,f_j^\dagger\}=\delta_{ij}$  (not
at this point to be associated with cuprate quasiparticles) living on
the sites of the original square lattice.  One may formally define
Jordan-Wigner hard-core bosons by introducing a ``string''
$\hat{S}$ winding
around the lattice,
\begin{equation}
  f_{\bf r} = b_{\bf r} \hat{S}_{\bf r},
\end{equation}
where 
\begin{equation}
  \hat{S}_{x+\frac{1}{2},y+\frac{1}{2}} = \prod_{x'<x} (-1)^{n_{x'+\frac{1}{2},y+\frac{1}{2}}}
  \prod_{y'>y}\prod_{x'} (-1)^{n_{x'+\frac{1}{2},y'+\frac{1}{2}}}.
\end{equation}
With this definition, the boson operators obey proper canonical
commutation relations at different sites, and have a hard-core
interaction on the same site.  For an ordinary dynamics, this
extremely non-local string presents unsolvable difficulties for
analytic treatment.  However, using the exact plaquette duality
appropriate for ring dynamics, one finds that one can write
\begin{equation}
\hat{S}_{x+\frac{1}{2},y+\frac{1}{2}} = e^{i(\theta_{x,y+1}-\theta_{x,y})}.
\end{equation}
Here we have ignored any possible boundary terms.  Up to this proviso,
the string becomes local in the $\theta$ variables!  Thus, in a rotor
representation with $b=e^{i\phi}$ (this requires a large $U n(n-1)$
term to maintain the hard-core constraint), one has
\begin{equation} 
  f_{x+\frac{1}{2},y+\frac{1}{2}} =
  e^{i\phi_{x+\frac{1}{2},y+\frac{1}{2}}} e^{i(\theta_{x,y+1}-\theta_{x,y})},
\end{equation}
which is directly analogous to the bosonization formula for spinless
fermions in one dimension!  This approach may lead to an understanding
of novel non-Fermi liquid states in fermionic ring models.

Clearly there are a range of applications to be explored based on the
present work.  We hope that some of these may ultimately enhance our
understanding of experimentally accessible strongly correlated materials.

\begin{acknowledgments}

  We are grateful to Anders Sandvik, Doug Scalapino, and T. Senthil for
  sparkling conversations.  L.B. and A.P. were supported by NSF grant
  DMR-9985255, and the Sloan and Packard foundations, M.P.A.F. by NSF
  Grants DMR-9704005 and PHY-9907949. \par

\end{acknowledgments}

\appendix
\section{Trotter decomposition}
\label{app:trotter}

The usual Trotter decomposition of $Z_\Box$ gives
\begin{equation}
  Z = \sum_{\{ \theta_{{\bf r}\tau} \}} \prod_{\tau=0}^{\tau=\beta}
  \langle \theta_{{\bf r}\tau+\epsilon} | e^{\epsilon K \sum_{\bf r}
    \cos(\pi N_{\bf r})} e^{-{\epsilon U \over
  2\pi^2} \sum_{\bf r} [\Delta_{xy} \theta_{\bf r}
- \pi\bar{n} ]^2  } |\theta_{{\bf r}\tau}\rangle,
\end{equation}
where we have used $\epsilon \ll 1$ to separate the imaginary-time
evolution operator into to factors.  The second can be directly
evaluated to give
\begin{equation}
  Z = \sum_{\{ \theta_{{\bf r}\tau} \}} \prod_{\tau=0}^{\tau=\beta} e^{-{\epsilon U \over
  2\pi^2} \sum_{\bf r} [\Delta_{xy} \theta_{{\bf r}\tau}
- \pi\bar{n} ]^2  }
  \prod_{\bf r}\langle \theta_{{\bf r}\tau+\epsilon} | e^{\epsilon K
    \cos(\pi N_{\bf r})}  |\theta_{{\bf r}\tau}\rangle.
\end{equation}
The latter matrix element can be written
\begin{eqnarray}
  \langle \theta' | e^{\epsilon K
    \cos(\pi N_{\bf r})}  |\theta\rangle & = & \langle
  \theta' | 1+ \epsilon K
    \cos(\pi N_{\bf r})  |\theta\rangle \nonumber \\
    & = & \delta_{\theta',\theta}
    + \frac{\epsilon K}{2} \left(\delta_{\theta',\theta+\pi}+
      \delta_{\theta',\theta-\pi}\right) \nonumber \\
    & = & \exp \left[ \frac{1}{\pi^2} \ln \left(\frac{\epsilon
          K}{2}\right) \left( \theta'-\theta\right)^2 \right],
\end{eqnarray}
correct to $O(\epsilon K)$.  This leads directly to
Eq.~(\ref{eq:Zdiscrete})

\section{Asymptotics\label{sec:asymptotics}}
In this appendix we calculate the asymptotics of the correlator
$c_{xy}(\tau,\tau_1,\tau_2)$, introduced in Sec.~\ref{sec:bosemetal}.
Using the general rules of Gaussian theories, one has
\begin{widetext}
\begin{eqnarray}
  c_{x,y}(\tau,\tau_1,\tau_2) & = & 2\left\langle \left(
      \vartheta_{xy}(\tau)+\vartheta_{00}(0)-
      \vartheta_{x0}(\tau_1)-
      \vartheta_{0y}(\tau_2)\right)^2\right\rangle_0  \nonumber \\   
  && \hspace{-1.0in} =  \!2\pi^2{\rm Re}\!\int_{\bf k}{\cal K}({\bf
    k})\frac{2\!+\!\left(e^{-E_{\bf 
        k}|\tau|}+e^{-E_{\bf k}|\tau_1-\tau_2|}\right)e^{-i(k_x x+k_y
    y)}
  \!-\!\left(e^{-E_{\bf k}|\tau_1|}+e^{-E_{\bf
        k}|\tau-\tau_2|}\right)e^{-ik_x x}
  \!-\!\left(e^{-E_{\bf k}|\tau_2|}+e^{-E_{\bf
        k}|\tau-\tau_1|}\right)e^{-i k_y y}}{E_{\bf k}}. 
\end{eqnarray}

The above integral was considered earlier for the special case
$\tau=\tau_1=\tau_2=0$, for which it was shown to grow logarithmically
with $x$ at fixed $y$ (and vice versa).  This growth {\sl with $x$ and
  $y$} persists when $\Omega_0|\tau|,\Omega_0|\tau_1|,\Omega_0|\tau_2|
\ll |x|$ (or $|y|$).  To see this, we differentiate
\begin{eqnarray}
  \label{eq:dcxy}
  \partial_x c_{xy} & = & 2\pi^2{\rm Re}\int_{\bf k} \frac{ik_x
    e^{-ik_x x}{\cal K}({\bf k})}{E_{\bf k}}\left[ e^{-E_{\bf
        k}|\tau_1|}+ e^{-E_{\bf k}|\tau-\tau_2|} - e^{-ik_y y}
    \left(e^{-E_{\bf k}|\tau|}+e^{-E_{\bf k}|\tau_1-\tau_2|}\right)\right] \nonumber \\ 
  &&\hspace{-0.8in} \sim 2 \int_0^{\pi}\!dk_y\, \frac{{\cal
      K}(0,k_y)}{\omega(k_y)} 
  \left[
    \frac{x}{x^2+\omega^2(k_y)\tau_1^2}+\frac{x}{x^2+\omega^2(k_y)(\tau\!-\!\tau_2)^2} - \cos k_y y \left( \frac{x}{x^2+\omega^2(k_y)\tau^2}+\frac{x}{x^2+\omega^2(k_y)(\tau_1\!-\!\tau_2)^2}\right)\right],
\end{eqnarray}
\end{widetext}
where $\omega(k_y)=\Omega(0,k_y)|2\sin(k_y/2)|$, and in the second
line we assumed (as valid for large $|x|$) the $k_x$ integral is
dominated by small $k_x$ and integrated it in that regime.  Hence for
$|x| \gg \Omega_0|\tau|_1, \Omega_0|\tau|$, one has
\begin{equation}
    c_{xy} \sim 4{\eta_v}(y)\ln|x| + f(\tau,\tau_1,\tau_2,y), \label{eq:cxy1}
\end{equation}
where $f$ is an unknown (but $x$-independent) function of $y$ and the imaginary
times.   We expect that $c_{xy}$ should cross over to a $\ln^2 |\Omega_0\tau_i|$
behavior if any of the $\Omega_0|\tau_i|$ become large compared to the
spatial distances involved.  However, it is unclear without further
calculation whether this crossover occurs when the imaginary time
argument becomes large compared to the {\sl smaller} (e.g. $y$ above)
or the {\sl larger} (e.g. $x$ above) of the two spatial coordinates.
To ascertain this information, we differentiate with respect to the
largest time (which we choose positive without loss of generality),
e.g. for $\tau>|\tau_1|,|\tau_2|$, 
\begin{eqnarray}
  \partial_\tau c_{xy} & = & 2\pi^2\int_{\bf k} {\cal K}({\bf k})\bigg[
    e^{-E_{\bf k}(\tau-\tau_1)-ik_y y} \nonumber \\
    & & + e^{-E_{\bf
        k}(\tau-\tau_2)-ik_x x}-e^{-E_{\bf k}\tau-i(k_x x+k_y 
      y)}\bigg] \nonumber \\
    & \approx & 2\pi^2\int_{\bf k} {\cal K}({\bf k})
    e^{-E_{\bf k}(\tau-\tau_1)}e^{-ik_y y},
\end{eqnarray}
where in last line we have kept the dominant term for $|x| \gg |y|,\Omega_0\tau$.  If we assume
$|y|\sim O(1)$, then for $\Omega_0(\tau-\tau_1) \gg 1$, the integral is dominated 
by ${\cal k}\approx 0$ (due to the logarithmic singularity in the
collective density of states), and one may approximate $e^{-ik_y
  y}\approx 1$, leading to $\partial_\tau c_{xy} \sim 2\sqrt{{\cal
    K}_0/{\cal U}_0}\ln|\Omega_0(\tau-\tau_1)|/(\tau-\tau_1)$, whence $f \sim \sqrt{{\cal
    K}_0/{\cal U}_0}\ln^2|\Omega_0(\tau-\tau_1)|$.  More generally, one can show 
that this behavior obtains provided $\Omega_0(\tau-\tau_1) \gg |y|$.
By identical arguments, $f$ also grows like $ \sqrt{{\cal
    K}_0/{\cal U}_0}\ln^2|\Omega_0\tau_2|$ for $\Omega_0|\tau_2| \gg
|y|$.  Thus, the integral defining $C_{xy}^{(4)}(\tau)$ is dominated
by $\tau_1 \approx \tau$ and $\tau_2 \approx 0$.  
Hence, we can focus on this case, and consider the behavior in general 
as a function of $|x| \gg 1$ and $\tau \gg 1$, but with no particular
relation between them.  One finds
\begin{eqnarray}
  c_{xy}(\tau,\tau,0) & \!\!\approx & \!\!4\pi^2 {\rm Re} \int_{\bf k} \frac{{\cal 
      K}({\bf k})}{E_{\bf k}}\left(1\!-\!e^{-ik_y
      y}\right)(1\!-\!e^{-E_{\bf k}|\tau|-ik_x x}). \nonumber
\end{eqnarray}
It is straightforward to see that this integral recovers the result,
Eq.~(\ref{eq:cxy1}) for $|x| \gg |\tau|$, and conversely,
$c_{xy}(\tau,\tau,0)\sim 4{\eta_v}(y) \ln\tau$ for $|\tau|\gg |x|$.
Putting all the above analysis together, one arrives at the general
approximate result of Eq.~(\ref{eq:cxyasympt}) in Sec.~\ref{sec:bosemetal}.

\section{Relevance of hopping}
\label{sec:relevance}

In this Appendix we compute the corrections to
the Bose metal imaginary-time equal-space correlator,
\begin{equation}
G(\tau) = \langle e^{i\varphi_{\bf r}(\tau)} e^{-i\varphi_{\bf r}(0)} \rangle ,
\end{equation}
to leading (quadratic) order in the Boson hopping term:
\begin{equation}
{\cal S}_t = - t_y \int_{\tau} \sum_{\bf r} 
\cos(\varphi_{{\bf r} + \hat{y}} - \varphi_{\bf r}) .
\end{equation}
Perturbing around the Gaussian Bose metal theory gives,
$G(\tau) = G_0(\tau) + G_2(\tau) + O(t_x^4)$, with
\begin{equation}
G_2(\tau) = {1 \over 2} \langle  e^{i\phi_{\bf r}(\tau)-i\phi_{\bf r}(0)}
{\cal S}_t^2 \rangle_{0,c}  ,
\end{equation}
where the connected correlator is taken with respect to the
Gaussian action.  Using Wick's theorem this can be re-expressed as,
$G_2(\tau) = t_x^2 G_0(\tau)  I(\tau)$, with
\begin{equation}
I(\tau) = \sum_{x_1,x_2,y_1} \int_{\tau_1,\tau_2} {\cal D}_0(x_{12},y_{12}) [e^{-A} -1]  ,
\end{equation}
where the function $A=A(x_1,x_2,y_1,\tau_1,\tau_2,\tau)$ is given by,
\begin{equation}
A= \int_{{\bf k},\omega}
e^{i{\bf k}\cdot{\bf r}_1 -i \omega \tau_1} B({\bf k},\omega,\tau)
(1 - e^{-ik_x x_{12} + i\omega \tau_{12}}) + h.c.  ,
\end{equation}

\begin{equation}
B({\bf k},\omega,\tau) = {\cal U}({\bf k})  { {(e^{i\omega\tau} -1)(1-e^{ik_y}) }
\over {\omega^2
+E_{\bf k}^2   }  }    .
\end{equation}
Here, $x_{12} = x_1-x_2$, $\tau_{12} = \tau_1 - \tau_2$
and ${\cal D}_0(x,\tau) \sim (x^2 + \tau^2)^{-\Delta}$ is the two-point
correlator for the tunneling operator evaluated
in the Bose metal.  To isolate the $\tau$ dependence of $I(\tau)$,
we re-scale $x_1,x_2,\tau_1,\tau_2$ by $\tau$, and $k_x$ and $\omega$
by $1/\tau$ (leaving $y$ and $k_y$ alone).  For large $\tau$ this gives,
$I(\tau) \sim \tau^{2(2-\Delta)} I(1)$, or equivalently,
\begin{equation}
{  {  G_2(\tau)  }  \over  { G_0(\tau) } } \sim
t_y^2 I(1) \tau^{2(2-\Delta)}   .
\end{equation}
We thus conclude that provided $I(1)$ is finite,
the large $\tau$ behavior of the Bose metal correlator,
$G_0(\tau)$, is unmodified if the scaling dimension
$\Delta >2$.  To show that $I(1)$ is finite, it suffices
to expand the exponential to second order in $A$, which gives:
\begin{equation}
I(1) = 8\int_{{\bf k},\omega} {\cal U}({\bf k}){\cal D}(k_x,\omega) {
  {(1-\cos(\omega))(1-\cos(k_y))  } 
\over   { (\omega^2 + E_{\bf k}^2  )^2   }   }   ,
\label{I1}
\end{equation}
with
\begin{equation}
{\cal D}(k_x,\omega) = \int_{x,\tau} {\cal D}_0(x,\tau)(1-e^{i(k_xx-\omega\tau)})  .
\end{equation}
Provided $\Delta>1$ the integrals over $x$ and $\tau$
converge, giving ${\cal D}(k_x,\omega) \sim (k_x^2 + \omega^2)^{\Delta-1}$.
Inserting this into Eqn.~\ref{I1}, one readily sees that the
${\bf k}$ and $\omega$ integrals are likewise convergent,
confirming that $I(1)$ is finite whenever $\Delta>1$.

\section{Z2 Gauge Theory and Ring Exchange\label{app:z2}}

\label{sec:Z2}

A recent formulation\cite{Z2A}\ of interacting electrons in
two dimensions has been developed which re-expresses the electron
operator in terms of ``spinons'' and ``chargons'' which are minimally
coupled to a $Z_2$ gauge field.  As with earlier $U(1)$ and $SU(2)$
gauge theory formulations, the ``spinons'' are taken as Fermions
carrying the spin of the electron but are electrically neutral.  The
bosonic ``chargons'' carry the electrons charge.  The $Z_2$ gauge theory
provides a convenient phenomenology for describing a fractionalized phase in
which the spinons and chargons are deconfined, and live as
well-defined particle excitations.  The deconfined phase is most
readily accessed by increasing the strength of a term in the gauge
theory Hamiltonian (with coefficient $K$, below) which suppresses the
fluctuations in the gauge field.  When the gauge theory is deep within
it's confined phase (with $K=0$), on the other hand, the gauge fields
can be formally integrated out and one recovers a Hamiltonian
expressible in terms of electron operators and composites built from
the electron such as the spin operator, together with a Cooper pair
field.  Here, we show that upon integrating out the gauge field with
small but non-zero $K$, one generates ring-exchange terms in the
effective Hamiltonian.  In the spin-sector of the theory, the leading
order ring term involves four-spin operators around an elementary
(square) plaquette, and is the familiar quartic form shown explicitly
below.  But there are also ring terms generated in the charge sector,
and the dominant one takes precisely the form in Eq. (1), where two
Cooper pairs are destroyed on opposite corners of an elementary square
plaquette, hopping to the other two corners.

Consider then a Hamiltonian version of the $Z_2$ gauge theory:
\begin{eqnarray}
H & = & H_c + H_{\sigma} + H_s, \\
H_c & \!\!\!= & \!\!\!-t_c\sum_{\langle{\bf r}{\bf r}^\prime\rangle}
\sigma^z_{{\bf r}{\bf r}^\prime}\! \left(b^{\dagger}_{\bf r} b_{{\bf
      r}^\prime} 
+ {\rm h.c.} \right) \nonumber\\
&+& U \sum_{\bf r} \left( b_{\bf r}^\dagger b_{\bf r} - 2 \bar{n} \right)^2, \\
H_{\sigma} & = & -K\sum_{\Box} \prod_{\Box} \sigma^z_{{\bf r}{\bf r}^\prime} - h \sum_{<{\bf r}j>} \sigma^x_{{\bf r}{\bf r}^\prime}, \\
H_s & = & -\sum_{\langle{\bf r}{\bf r}^\prime\rangle} \sigma^z_{{\bf r}{\bf r}^\prime}\left[t_s \left(f^{\dagger}_{\bf r} f_{{\bf r}^\prime} + 
{\rm h.c.} \right) \right. \nonumber \\
 &+& \Delta_{{\bf r}{\bf r}^\prime} \left. \left(f_{{\bf r}\uparrow}
     f_{{\bf r}^\prime\downarrow} - f_{{\bf r}^\prime \downarrow}
     f_{{\bf r} \uparrow}  + {\rm h.c.} \right) \right]
,
\end{eqnarray} 
where ${\bf r}$ and ${\bf r}^\prime$ denote sights of a 2d square
lattice.  Here $b^{\dagger}_{\bf r}$ creates a chargon at site ${\bf
  r}$ while $f^{\dagger}_{{\bf r}\alpha}$ creates a spinon with spin
$\alpha = \uparrow, \downarrow$ at site ${\bf r}$. The operator
$b^{\dagger}_{\bf r}b_{\bf r}$ measures the number of charge $e$
bosons at site ${\bf r}$, and the Hubbard-like $U$ term sets the mean
Cooper pair density, $n_{\bf r} \equiv  b^{\dagger}_{\bf r}b_{\bf
  r}/2$, to be $\bar{n}$.  With this convention, the half-filled
electron model with charge $e$ per site corresponds to $\bar{n} =
1/2$.  The constant $\Delta_{{\bf r}{\bf r}^\prime}$ contains the
information about the pairing symmetry of the spinons. The
$\sigma^z_{{\bf r}{\bf r}^\prime}, \sigma^x_{{\bf r}{\bf r}^\prime}$
are Pauli spin matrices that are defined on the links of the lattice.
The full Hamiltonian is invariant under the $Z_2$ gauge transformation
$b_{\bf r} \rightarrow -b_{\bf r}, f_{\bf r} \rightarrow -f_{\bf r}$
at any site ${\bf r}$ of the lattice accompanied by letting
$\sigma^z_{{\bf r}{\bf r}^\prime} \rightarrow -\sigma^z_{{\bf r}{\bf
    r}^\prime}$ on all the links connected to that site.  This
Hamiltonian must be supplemented with the constraint equation
\begin{equation}
  {\cal G}_{\bf r} = \prod_{{\bf r}^\prime \in {\bf r}}\sigma^x_{{\bf
      r}{\bf r}^\prime} e^{i\pi \left(f^\dagger_{\bf r} f^{\vphantom\dagger}_{\bf r} +
      b^\dagger_{\bf r}b^{\vphantom\dagger}_{\bf r} \right)} = 1. 
\end{equation}
Here the product over $\sigma^x_{{\bf r}{\bf r}^\prime}$ is over all
links that emanate from site ${\bf r}$.  The operator ${\cal G}_{\bf
  r}$, which commutes with the full Hamiltonian, is the generator of
the local $Z_2$ gauge symmetry. Thus the constraint ${\cal G}_{\bf r}
=1$ simply expresses the condition that the physical states in the
Hilbert space are those that are gauge invariant.

Generally, the gauge field dynamics is very complicated, but it
simplifies considerably when either $h$ or $K$ greatly exceeds the
strength of the couplings to the matter fields, $t_c, t_s$ and
$\Delta_{{\bf r}{\bf r}^\prime}$.  For example, when $K \rightarrow
\infty$, fluctuations in $\sigma^z_{{\bf r}{\bf r}^\prime}$ are
suppressed completely, and one can choose a gauge with the
$\sigma^z_{{\bf r}{\bf r}^\prime}=1$ on every link.  This corresponds
to a deconfined phase in which the spinons and chargons exist as bona
fide particle excitations.  On the other hand, when $h \rightarrow
\infty$ one has $\sigma^x_{{\bf r}{\bf r}^\prime} \approx 1$, and it
is convenient to integrate out the gauge field perturbatively.
Focusing for simplicity on the Mott insulator at half-filling, and in
the latter large $h$ limit, with $\bar{n} = 1/2$ with $U \rightarrow
\infty$, at lowest order in $t_s$ and $\Delta$ one recovers the
Heisenberg spin-model,
\begin{equation}
H_2 = J_2 \sum_{<{\bf r}{\bf r}^\prime>} {\bf S}_{\bf r} \cdot {\bf S}_{{\bf r}^\prime}   ,
\end{equation}
with $J_2 = (t_s^2 + \Delta^2)/h$.  Here, ${\bf S}_{\bf r} \equiv
\frac{1}{2} f^\dagger_{{\bf r} \alpha} {\bf s}_{\alpha \beta} f_{{\bf
    r} \beta}$, with ${\bf s}$ a vector of Pauli spin matrices.  Note
that in the $h,U \rightarrow \infty$ limit the gauge constraint
becomes a single-occupancy constraint: $e^{i \pi f^\dagger_{\bf r}
  f^{\vphantom\dagger}_{\bf r}} =1$.  Upon inclusion of a small
non-zero coupling $K$, one generates the spin ring exchange term in
Eq.~(\ref{eq:plaqspin}), upon integrating out the gauge field, with
$J^s_\Box = K (t_s^4 + \Delta^4)/(h+U)^4$.  Notice that the strength
of the 4-spin ring exchange interaction, $J^s_\Box$, is proportional
to the gauge theory coupling $K$.  This suggests that
spin models with appreciable ring exchange interactions are good
candidate models to exhibit fractionalized phases.

It is very instructive to examine the charge sector of the
theory one generates
upon integrating out the gauge field by similarly expanding perturbatively
in small $t_c$.  Generally, one will generate
various Cooper pair hopping processes.   To leading 
order in $K$ the Cooper pair plaquette ring term of
Eq.~(\ref{eq:plaqcharge}) is generated, with
$J^c_\Box = K t_c^4/(h+U)^4$.

\section{RK Manipulations}

\label{sec:saddle}

We begin by inserting
the integral representations of $P_{x/y}$ and the explicit
representation of $P_{\rm flip}$, inherited from
Eqs.~(\ref{eq:projectors}) and Eq.~(\ref{eq:Pflip}), respectively.
The sums over $\{\sigma_{xy}\}$ can be explicitly performed, and we
find
\begin{equation}
  \left\langle P_{\rm
      flip}(x,y;\{\sigma\})\right\rangle_{\{\sigma_{xy}\}} = \langle
  F_{xy}[\{\theta,\phi\}] \rangle_{\{\theta_x,\phi_y\}},
  \label{eq:ptof}
\end{equation}
where
\begin{equation}
  \langle {\cal O}\rangle_{\{\theta_x,\phi_y\}} =
  \frac{1}{Z_{\theta\phi}} \prod_{a=1}^L\int_0^{2\pi}
  \left(\frac{d\theta_a}{2\pi}
    \frac{d\phi_a}{2\pi}\right) {\cal O} e^{-s[\{\theta_x,\phi_y\}]}.
  \label{eq:expect}
\end{equation}
In these expressions, the transformed ``flippability'' function is
\begin{equation}
  F_{xy}[\{\theta,\phi\}] = \frac{1}{8}\prod_{u,v=0,1} \sec
  (\theta_{x+u}-\phi_{y+v}),
\end{equation}
and the classical ``action'' is given by
\begin{eqnarray}
  s[\{\theta_x,\phi_y\}] & = & -\sum_{x,y=1}^L \ln \cos(\theta_x-\phi_y) +
  i m L \sum_{a=1}^L (\theta_a-\phi_a) \nonumber \\
  & & + \frac{M^2}{2}\big(\sum_a
    \theta_a -\phi_a\big)^2. \label{eq:bareaction}
\end{eqnarray}
In Eq.~(\ref{eq:bareaction}), we had added the final $M^2$ term to fix
a redundancy in the description.  In it's absence, the action $s$
possesses a continuous translational symmetry inherited from the
definition of the projection operators, $\theta_x \rightarrow
\theta_x+\lambda$, $\phi_y \rightarrow \phi_y + \lambda$, for all
$x,y$.  This represents a redundancy in the constraints due to the
fact that the total particle number of the system is given both by the
sum of the row particle numbers and by the sum of the column particle
numbers.  The addition of the $M^2$ term has no effect on physical
quantities, and has the benefit of lifting the unnecessary zero-mode
of $s$.  We will in fact take $M^2\rightarrow \infty$ ultimately for
simplicity at an appropriate stage of the calculation.

Note the $L$ prefactor in front of the second term in
Eq.~(\ref{eq:bareaction}), which suggests a saddle point approximation
for large $L$ (in the first term, note also that there are $L^2$
components in the sum but only $2L$ variables).  The saddle point
conditions, $\frac{\partial s}{\partial\theta_x} = \frac{\partial
  s}{\partial\phi_y} = 0$,  are
\begin{eqnarray}
   \sum_{y=1}^L
  \tan(\theta_x-\phi_y) & = & -i m L - M^2 \sum_{a} (\theta_a+\phi_a), \\
  \sum_{x=1}^L
  \tan(\theta_x-\phi_y) & = & -i m L - M^2 \sum_{a} (\theta_a+\phi_a).
\end{eqnarray}
These are solved by the uniform (imaginary) solution $\theta_x = - \phi_y = -i
\eta/2$, with
\begin{equation}
  \tanh \eta=m.
\end{equation}
Next, expanding around this saddle, we let $\theta_x=-i \eta/2 +
\frac{1}{2}(\psi_{+x} + \psi_{-x})$, $\phi_y = i \eta/2 +
\frac{1}{2}(\psi_{+x} - \psi_{-x})$.  Expanding the action to cubic
order in $\psi_\pm$, as justified by the following analysis, one finds
$s=s_0 + s_2 + s_3$, with
\begin{eqnarray}
  s_0 & = & \frac{1}{2}L^2 \left( m \ln
    \left[\frac{1+m}{1-m}\right]+\ln(1-m^2)\right), \\
  s_2 & = & \frac{1}{4} (1-m^2) \bigg[ L \sum_{a} \label{eq:s0def}
    \left(\psi_{+a}^2+\psi_{-a}^2\right) \nonumber \\
    & &  +\sum_{a,b} \left( (2M^2-1)\psi_{+a}\psi_{+b} +
      \psi_{-a}\psi_{-b}\right) \bigg], \label{eq:s2def} \\
  s_3 & = & -\frac{i}{12} m (1-m^2) \bigg[L \sum_a
    \left(\psi_{-a}^3+3\psi_{+a}^2\psi_{-a}\right)  \nonumber \\
    & & \hspace{-0.4in}+
    3\sum_{a,b}\left(\psi_{-a}^2 \psi_{-b}
    + \psi_{+a}^2 \psi_{-b} - 2\psi_{+a} \psi_{+b} \psi_{-b}\right)
  \bigg]. \label{eq:s3def}
\end{eqnarray}
The constant term $s_0$ drops out of the observables in which we are
interested.  The quadratic action, $s_2$, governs small fluctuations
of $\psi_\pm$, treating treating cubic and higher
order terms as perturbations.  The resulting
Gaussian averages obey Wick's theorem, governed by the propagators that are
obtained in the usual way by inverting the quadratic form in $s_2$.
\begin{eqnarray}
  G_{ab}^+ = \langle \psi_{+a}\psi_{+b}\rangle_{s_2} & = &
  \frac{2}{1-m^2}\left(\frac{\delta_{ab}}{L}-\frac{1}{L^2}\right),
  \label{eq:psi+}\\
  G_{ab}^- = \langle \psi_{-a}\psi_{-b}\rangle_{s_2} & = &
  \frac{2}{1-m^2}\left(\frac{\delta_{ab}}{L}-\frac{1}{2L^2}\right),
  \label{eq:psi-}
\end{eqnarray}
where in Eq.~(\ref{eq:psi+}) we have taken for simplicity the
$M^2\rightarrow\infty$ limit.  Eqs.~(\ref{eq:psi+}-\ref{eq:psi-})
imply that the fluctuations (variance) of individual $\psi_{a\pm}$
fields are small ($O(1/L)$), and moreover the correlations between
fields at different $a\neq b$ are even smaller ($O(1/L^2)$).

For this reason, the energy density $\epsilon(m)$ is determined by the
saddle-point value alone, i.e.
\begin{equation}
 \left\langle P_{\rm
     flip}(x,y;\{\sigma\})\right\rangle_{\{\sigma_{xy}\}} =
 F_{xy}[\theta=-\phi=-i\eta/2],
\end{equation}
leading directly to Eq.~(\ref{eq:energydensity})

The ``single-particle'' gap $\Delta_1$ is slightly more involved,
since we need the energy up to terms of order $1/L$.  To
calculate it, we require $E_1(m;\lambda)$, which is obtained from the
expectation value of the flippability in a state with $\lambda$
additional bosons on row and column 1.  This is obtained, according to
Eq.~(\ref{eq:extraproj}), by slightly modifying the projection
operators on this row and column.  This amounts to adding an additional
source term to the action, $s[\{\theta,\phi\}] \rightarrow s[\{\theta,\phi\}] +
2i\lambda(\theta_1-\phi_1)$, but otherwise calculating the same
expectation value as in Eq.~(\ref{eq:ptof}).  Transforming to the
$\psi_\pm$ variables as before, this additional linear term can be
removed from the action by shifting $\psi_{a-} \rightarrow \psi_{a-} -
2 i \lambda G_{a1}^-$, which ``completes the square'' in the action
using $s_2$ from Eq.~(\ref{eq:s2def}).  Dependence upon $\lambda$ thereby
moves into $F$ and $s_3$:
\begin{widetext}
\begin{eqnarray}
E_1(m;\lambda) &   = &
-v \sum_{xy} \left\langle
 F_{xy}[\theta_a=-\frac{i}{2}\eta+ \frac{1}{2}(\psi_{+a}+\psi_{-a}
 -2i\lambda G_{a1}^-), \phi_a = \frac{i}{2}\eta+ \frac{1}{2}(\psi_{+a}-\psi_{-a}
 +2i\lambda G_{a1}^-) ]\right\rangle^\prime, \label{eq:nasty}
\end{eqnarray}
\end{widetext}
where the prime indicates the expectation value is as defined in
Eq.~(\ref{eq:expect}), except that $s_3[\psi_{+a},\psi_{-a}]$ is
replaced by $s'_3[\psi_{+a},\psi_{-a}]=
s_3[\psi_{+a},\psi_{-a}-2 i \lambda G_{a1}^-]$.
This can be evaluated by expanding $F_{xy}$ above in $\psi_\pm$ and
$\lambda$ (the latter since $G_{a1}^- \ll 1$), and further evaluating
the expectation value perturbatively in $s_3$.  A careful examination
of these expansions shows that there are only two contributions to
$\Delta_1$.  The first comes at $O(\psi_\pm^0)$ from the pure
saddle-point contribution to $F_{xy}$, expanded to second order in
$\lambda$.  The second comes from expanding $F_{xy}$ to
$O(\psi_\pm^2)$ at $\lambda=0$ and evaluating the expectation value in
Eq.~(\ref{eq:nasty}) to first order in $s_3$ (itself expanded to
$O(\lambda^2 \psi_\pm)$).  Adding the two gives the result in
Eq.~(\ref{eq:opg}).



\begin{thebibliography}{99}
\bibitem{bosecondensenote} The crossover to the superfluid state is
  slightly subtle if for some reason the most relevant tunneling
  operator has $m>1$.  In this case, these strongest tunneling processes
  conserve the boson number on $m$ subsets of the rows and $m$ subsets 
  of the columns, leading to a state which should be thought of as a
  collection of $m^2$ interpenetrating and coexisting superfluids
  (i.e. with $m^2$ second sound modes).  However, this state is in
  turn unstable to arbitrarily weak tunneling $t_{11}$, which
  generates a Josephson coupling between these superfluids.  This
  leads to a single (fully mixed) second sound mode at low energies.
\end{thebibliography}
\end{document}